\documentclass[aps,amsmath,amssymb,amsfonts,oneside,12pt]{revtex4}
\usepackage{graphicx}
\usepackage{wasysym}
\newcommand{\bm}{\boldmath}
\newcommand{\D}{\displaystyle}
\newcommand{\vek}[1]{\mbox{\bm ${#1}$}}
\usepackage{arcs}
\usepackage{extraipa}
\usepackage{textcomp}
\usepackage{tipa}
\usepackage{pdflscape}
\usepackage{longtable}
\usepackage{booktabs}

\begin{document}

\thispagestyle{empty}
\renewcommand{\theequation}{\arabic{section}.\arabic{equation}}
\vspace{3cm}
\begin{center}{\large\bf The two-body reduced density matrix of the high-density electron gas}
\end{center}

\begin{center}{\sc P. Ziesche} \\

{Max-Planck-Institut f\"ur Physik komplexer Systeme \\
N\"othnitzer Str. 38, D-01187 Dresden, Germany}, pz@pks.mpg.de\\
\end{center}
\begin{center}
\end{center}
\vspace{-2.5cm}
\date{draft of \today}
\vspace{2cm}
\begin{abstract}
\noindent
For the first time, the {\it cumulant} 2-body reduced density matrix (= 2-matrix) of the spin-unpolarized homogeneous electron gas (HEG) is considered. This $\gamma_{\rm c}$ proves to be the common source
for both the momentum distribution $n(k)$ and the static structure factor $S(q)$. Within many-body perturbation theory, this $\gamma_{\rm c}$ is given by only {\it linked} diagrams (with 2 open 
particle-hole lines as well as with closed loops and interaction lines). Here it is worked out in detail, how the 1-body 
quantity $n(k)$ follows from the 2-body quantity $\gamma_{\rm c}$ - through a certain contraction procedure, cf Eqs.(\ref{b25})-(\ref{b28}). In particular, this $\gamma_{\rm c}$ is developed 
for the high-density HEG. Its correctness is checked by deriving from it $n(k)$ and $S(q)$, known from the random-phase approximation (RPA). This study opens the way to a more sophisticated HEG description 
in terms of cumulant geminals or/and variational methods. Besides, the cumulant structure factor (CSF) of the exchange in lowest order is explicitly given and sum rules for the CSFs and their 
small- and large-$q$ behavior (beyond RPA) are systematically summarized within the plasmon sum rule, coalescing theorems, and the inflexion-point trajectory. 
\begin{center}
\underline{List of Symbols/Shorthands/Abbreviations}\\
\end{center}
RDM reduced density matrix, SR sum rule, $\rho$ homogeneous electron density \\
MD momentum distribution $n(k)$, $f(r)$ 1-body RDM\\
PD pair density $g(r)$, CPD cumulant PD $h(r)$: $g(r)=1-\frac{1}{2}f^2(r)-h(r)$\\
SF structure factor $S(q)$, CSF cumulant SF $C(q)$: $S(q)=1-\frac{1}{2}F(q)-C(q)$\\
"a" antiparallel spin: $g_{\rm a}(r)\leftrightarrow S_{\rm a}(q)$, $h_{\rm a}(r)\leftrightarrow C_{\rm a}(q)$ \\
"p" parallel spin: $g_{\rm p}(r)\leftrightarrow S_{\rm p}(q)$, $h_{\rm p}(r)\leftrightarrow C_{\rm p}(q)$ \\
FT Fourier transform, $F(q)$ = FT of $f^2(r)$, FS Fermi surface $|\vek k|=1$
\end{abstract}
\maketitle

\newpage

\section{Introduction}
\setcounter{equation}{0}
\noindent
The lowest-level quantum-kinematics of an extended many-electron system (within the Born-Oppenheimer approximation) is contained in 1- and 2-body densities and reduced density matrices (RDMs): $\rho_1(1)$= electron 
density, 
$\gamma_1(1|1')$= 1-body RDM (1-matrix), $\rho_2(1,2)$= pair density (PD), $\gamma_2(1|1',2|2')$= 2-matrix. The well-tried and widely used density functional theory (DFT) yields $\rho_1$ (but not $\gamma_1$ and 
$\rho_2$). If a density-matrix functional theory would exist, it would yield $\gamma_1$ (but not $\rho_2$); if a pair-density functional theory would exist, it would yield $\rho_2$ (but not $\gamma_1$). If an 
effective 2-body scheme would exist, which yields approximately the 2-matrix $\gamma_2$, then from it would follow both the PD $\rho_2$ (by taking the diagonal elements) and the 1-matrix $\gamma_1$ (by means of 
the contraction procedure $2'=2$ and $\int d2\ \gamma_2(1|1',2|2)$). For extended systems this requires the cumulant decomposition of $\gamma_2$. In this paper (it continues from \cite{Zie2,Zie13,Zie16} and is 
related to \cite{Ons,GGSB}), the {\bf cumulant 2-matrix} $\gamma_{\rm c}$ (dimensionless: $\chi$) is developed as a decisive key quantity and its properties are 
studied for a "simple" model system, namely the high-density spin-unpolarized homogeneous electron gas (HEG).  \\

\noindent
Although not present in the Periodic Table, this HEG is still an important and so far unsolved 
model system for the electronic structure theory, cf e.g. \cite{Tos}. Its advantage: pure "correlation" and no "multiple-scattering". In its spin-unpolarized version, the 
HEG ground state (GS) is characterized by only one parameter $r_s$, such that a sphere with this {\bf Wigner-Seitz radius} 
$r_s$ contains {\it on average} one electron \cite{Zie1}. It determines the Fermi wave number as 
$k_{\rm F}=1/(\alpha r_s)$ in atomic units (a.u.) with $\alpha =[4/(9\pi)]^{1/3}\approx 0.521062$ and it measures simultaneously 
both the interaction strength and the homogeneous electron density $\rho$, such that high density corresponds to weak interaction and 
hence weak correlation \cite{foo}. For recent papers on this limit cf \cite{Cio0,Zie2,Zie3,Mui,Cio,Zie4,Gla2,Zie5,Zie11}. $r_s=0$ corresponds to the ideal Fermi gas.
Its GS energy per particle is $e_0=3/10$, measured in units of $k_{\rm F}^2$. The Coulomb repulsion $v(q)=q_c^2/q^2$ with $q_c^2=4\alpha r_s/\pi$ causes deviations.
This is in lowest order the exchange energy $e_{\rm x}=-(3/4\pi)\alpha r_s$. But in next order the long range of the Coulomb repulsion makes the correlation energy 
$e_{\rm corr}$, defined by $e=e_0+e_{\rm x}+e_{\rm corr}+O(r_s)$, to behave non-analytically at the high-density limit: 
$e_{\rm corr}\sim r_s^2[\ln r_s+{\rm const}+O(r_s)]$ \cite{Hei,Ma,GB,Ons}. This non-analytical behavior of $e$ carries over to the kinetic and potential components, 
$t$ respectively $v$, through the virial theorem after March (1958)\cite{Mar}: 
\begin{equation}\label{a1}
v=r_s\frac{d}{d r_s}e\ ,\quad  t=-r_s^2\frac{d}{dr_s}\frac{1}{r_s}e\ . 
\end{equation}
The energy components $t$ and $v$ follow from the simplest quantum-kinematical quantities, namely $n(k)$ and $S(q)$, the 
momentum distribution (MD) and the (static) structure factor (SF), respectively:
\begin{equation}\label{a2}
t=\int\limits_0^\infty d(k^3)\ n(k)\frac{k^2}{2}\ ,\quad \int\limits_0^\infty d(k^3)\ n(k)=1 \ ,\quad \int\limits_0^\infty d(k^3)\ n(k)[1-n(k)]=c\ , 
\end{equation}
\begin{equation}\label{a3}
v=-\int\limits_0^\infty \frac{d(q^3)}{3\cdot 4}\ [1-S(q)]\frac{q_c^2}{q^2}\ ,\quad S(0)=0\ , \quad \int\limits_0^\infty d(q^3)\ [1-S(q)]=1+c_1\ .
\end{equation}
$k$ and $q$ are measured in units of $k_{\rm F}$ and $r$ in units of $k_{\rm F}^{-1}$.
The quantity $c$ is hereforth referred to as L\"owdin parameter, because L\"owdin was the first one who queried the
meaning of the trace of the squared 1-matrix. Because of $0\leq n(k)\leq 1$, $c$ measures the $r_s$-dependent non-idempotency of $n(k)$, being zero for $r_s=0$ (ideal Fermi gas) and increasing with $r_s$.
Thus $c$ measures simultaneously the strength of correlation. Note its particle-hole symmetry. $c_1$ is another (short-range) correlation parameter, also vanishing for $r_s\to 0$. \\ 

\noindent
Describing {\bf electron pairs}, one has to distinguish SFs $S_{\rm a,p}(q)$ for pairs with antiparallel spin ("a") and with parallel spins ("p") and corresponding pair densities (PDs) $g_{\rm a,p}(r)$, 
where $g_{\rm a}(r)$ and $g_{\rm p}(r)$ describe the Coulomb hole and the Fermi hole, respectively. The $g_{\rm a,p}(r)$ and $S_{\rm a,p}(q)$ are mutually related through Fourier transform (FT): 
\begin{eqnarray}\label{a4} 
g_{\rm a}(r)-1&=&\int\limits_0^\infty d(q^3)\frac{\sin qr}{qr}\ S_{\rm a}(q) \quad \leftrightarrow \quad S_{\rm a}(q)=
\alpha^3\int\limits_0^\infty d(r^3)\frac{\sin qr}{qr}\ \frac{1}{2}[g_{\rm a}(r)-1]\ , \\
g_{\rm p}(r)-1&=&\int\limits_0^\infty d(q^3)\frac{\sin qr}{qr}\ [S_{\rm p}(q)-1] \quad \leftrightarrow \quad S_{\rm p}(q)-1=
\alpha^3\int\limits_0^\infty d(r^3)\frac{\sin qr}{qr}\ \frac{1}{2}[g_{\rm p}(r)-1]\ . \nonumber 
\end{eqnarray}
From this follow the "total" PD $g(r)=[g_{\rm a}(r)+g_{\rm p}(r)]/2$ and the "total" SF $S(q)=S_{\rm a}(q)+S_{\rm p}(q)$:
\begin{equation}\label{a5}
g(r)-1=\int\limits_0^\infty d(q^3)\frac{\sin qr}{qr}\ \frac{1}{2}[S(q)-1] \quad \leftrightarrow \quad S(q)-1=
\alpha^3\int\limits_0^\infty d(r^3)\frac{\sin qr}{qr}\ [\ g(r)-1]\ .  \\
\end{equation} 
 $S_{\rm a,p}(0)=0$ fix the normalizations of $1-g_{\rm a,p}(r)$, known as the perfect screening SR, whereas the coalescing (or on-top) values $g_{\rm a}(0)$ and $g_{\rm p}(0)=0$ fix the normalizations of 
$S_{\rm a}(q)$ and $S_{\rm p}(q)-1$, respectively. $g_{\rm a}(0)=1-c_1$ or $g(0)=(1-c_1)/2$ shows the 
physical meaning of $c_1$. It determines the short-range correlation in terms of $g_{\rm a}(0)$, the on-top value of the Coulomb hole  $g_{\rm a}(r)$ with $0\leq g_{\rm a}(0)\leq 1$. For further details (normalizations) cf (\ref{B1}), (\ref{B2}).  \\

\noindent
In view of Eqs.(\ref{a2}) and (\ref{a3}), one may ask which peculiarities of $n(k)$ and $S_{\rm a,p}(q)$ cause the 
non-analyticities of $t$ and $v$, respectively. As shown in \cite{Zie3}, the above-mentioned 
drastic changes, when switching on the long-range Coulomb interaction, show up in the redistribution of the non-interacting MD 
$n_0(k)=\Theta (1-k)$ within thin layers (thickness $q_{\rm c}$) inside and outside the Fermi surface $|\vek k|=1$ and with a remaining finite jump discontinuity at 
$|\vek k|=1$:
$z_{\rm F}=n(1^-)-n(1^+)$, $0\leq z_{\rm F}\leq 1$. They show up also in the behavior of $S(q)$ within a small 
spherical region (radius $q_{\rm c}$) around the origin of the reciprocal space with the plasmon SR $S(q\ll q_{\rm c})=q^2/(2\omega_{\rm pl})+\cdots$ \cite{Pin,Thom}, which causes an inflexion point 
$q_{\rm infl}, S_{\rm infl}\sim 
\omega_{\rm pl}$, where $\omega_{\rm pl}=q_{\rm c}/\sqrt 3$ is the plasma frequency, measured in units of $k_{\rm F}^2$. The mentioned non-analyticities and 
redistributions let ordinary perturbation theory fail, e.g. with $S(q\to 0)\sim 1/q$ and $n(k)=n_0(k)+\Delta n(k)$ with $n_0(k)=\Theta (1-k)$ and $\Delta n(k\to 1^\pm)\sim \pm 1/(k-1)^2$. That the GS energy $e(r_s)$
diverges in 2nd order, has been shown by Heisenberg (1947) 
\cite{Hei}. Macke (1950) \cite{Ma} has repaired this failure by means of an appropriate, physically plausible {\bf partial summation of higher-order perturbation terms} for the GS
energy per particle $e(r_s)$, a way which has been developed further by Gell-Mann and Brueckner (1957) \cite{GB}. This so-called ring-diagram summation, cf Fig.1, is also known as the 
random phase approximation (RPA). This summation has been developed for $n(k)$ by Daniel/Vosko (1960) \cite{Da} and Kulik (1961) \cite{Ku}, an analytical extrapolation for $n(k)$ is given in 
\cite{Zie7,GGZ}, for the spin-polarized case see \cite{Zie8}, recent quantum Monte Carlo calculations of $n(k)$ for $r_s=1,\cdots, 10$ are in \cite{Holz}, $z_{\rm F}$ for high-densities is in \cite{Zie0}, 
$z_{\rm F}$ calculated for $r_s<55$ is in \cite{Nech}. The ring-diagram summation for $S(q)$ has been done by Glick/Ferrell (1960) \cite{Gli}, Geldart (1967) \cite{Gel2}, and Kimball (1976) \cite{Kim3}. 
For attempts to go beyond RPA cf eg  \cite{Holz,Cep}.    
The on-top behavior of $g_{\rm a,p}$ (describing the short-range correlation) influence the large-$q$ behavior of $S_{\rm a,p}(q)$ and $n(k)$. This comes from theorems referred here as 
the coalescing cusp and curvature theorems, cf \cite{Kim1} and \cite{Kim4} and App.B.   \\ 

\noindent
With $n(k)$ also $t$ is available (but not $v$) and with $S(q)$ it is available $v$ (but not $t$). Does a quantity exist which contains both $n(k)$ and $S(q)$? "Old" and recent attempts in that direction (with variational or direct calculations of the 2-matrix) are in \cite{You,Noo,Nak}. 
In \cite{Zie11} it has been shown that the self-energy $\Sigma (k,\omega)$ as a {\it functional} of
$t(\vek k)=k^2/2$ and $v(\vek q)=q_{\rm c}^2/q^2$ is such a quantity \cite{foo1}. Here it is shown - using the concept of reduced density matrices (RDMs) - that $n(k)$ and $S(q)$
have their common origin in a quantity called 'cumulant 2-matrix', $\chi(1|1',2|2')$ with the symbolic variable $1=(\vek r_1,\sigma_1)$. This matrix appears if one tries to represent the 2-matrix of an 
interacting system in terms of the 1-matrix (according to an independent-particle or Hartree-Fock model). There remains an irreducible part, which can not be reduced due to correlation.
For the PD and for the SF, these {\bf cumulant decompositions} are 
\begin{equation}\label{a6}
g_{\rm a}(r)=1-h_{\rm a}(r)\ ,  \hspace{1.9cm} S_{\rm a}(q)=-C_{\rm a}(q)\ , 
\end{equation}
\begin{equation}\label{a7}
g_{\rm p}(r)=1-f^2(r)-h_{\rm p}(r)\ ,  \quad S_{\rm p}(q)=1-\frac{1}{2}F(q)-C_{\rm p}(q)\ .  
\end{equation}
Note the asymmetry with respect to "a" (Coulomb hole) and "p" (Fermi hole).
With the "total" cumulants $h(r)=[h_{\rm a}(r)+h_{\rm p}(r)]/2$ and $C(q)=C_{\rm a}(q)+C_{\rm p}(q)$ it follows
\begin{equation}\label{a8}
g(r)=1-\frac{1}{2}f^2(r)-h(r)\ ,     \quad   S(q)=1-\frac{1}{2}F(q)-C(q)\ . 
\end{equation}
$f(r)$ is the dimensionless 1-matrix [following from $n(k)$, cf (2.4)] and $F(q)$ - referred to as HF-function - is the FT of $f^2(r)$, cf (\ref{B15}). $h_{\rm a,p}(r)$ are the cumulant PDs (CPDs) and their FTs are the cumulant SFs 
(CSFs) $C_{\rm a,p}(q)$:
\begin{equation}\label{a9}
h_{\rm a}(r)=\int\limits_0^\infty d(q^3)\frac{\sin qr}{qr}\ C_{\rm a}(q) \quad \leftrightarrow \quad C_{\rm a}(q)=
\alpha^3\int\limits_0^\infty d(r^3)\frac{\sin qr}{qr}\ \frac{1}{2}h_{\rm a}(r)\ , 
\end{equation}
\begin{equation}\label{a10}
h_{\rm p}(r)=\int\limits_0^\infty d(q^3)\frac{\sin qr}{qr}\ C_{\rm p}(q) \quad \leftrightarrow \quad C_{\rm p}(q)=
\alpha^3\int\limits_0^\infty d(r^3)\frac{\sin qr}{qr}\ \frac{1}{2}h_{\rm p}(r)\ .
\end{equation}
Within RDM theory one can show that the cumulant 2-matrix $\chi(1|1',2|2')$ is the source for: (i) the  CSF $C(q)$ [or equivalently the CPD $h(r)$] and (ii) also for the MD $n(k)$. Whereas (i) results simply from 
$\chi(1|1,2|2)$, i.e. taking the diagonal elements $1'=1$, $2'=2$, the case (ii) follows from the slightly more complicated RDM contraction of $\chi(1|1',2|2')$ with $2'=2$, 
$\int d2\ \chi(1|1',2|2)$ and FT. The Eqs.(\ref{a9}) and (\ref{a10}) contain the CSF properties (\ref{B1}) and (\ref{B2}) needed below. \\

\noindent
General remarks on {\bf RDMs and cumulants}: Whereas finite many-body systems can be described by many-body wave functions $\Phi(1,2,\cdots)$ as solutions of a Schr\"odinger 
equation, extended systems have to be described by a hierarchy of $N$-representable RDMs as solutions of the BBGKY-like hierarchy of contracted Schr\"odinger equations. And: whereas for
finite systems the concept of cumulant (2-body, 3-body, ...) matrices {\it can} be applied, for extended systems it is a {\it must}, because - unlike the RDMs - all these cumulant matrices are 
{\it size-extensive} entities, what allows the thermodynamic limit with $N\to \infty$, $\Omega\to \infty$, $\rho=N/\Omega={\rm const}$. Closely related with this property is that they 
are given within perturbation theory by non-vacuum {\it linked} Feynman diagrams. In App.A, the systematic definition of cumulant matrices in terms of generating functionals 
is summarized. Recent (quantum chemical) papers on RDMs, cumulants, contracted Schr\"odinger equations, correlation strength, correlation entropy, quantum entanglement, Berry phases etc. are \cite{Ko}-\cite{KuMu} 
and refs. therein. \cite{You} and \cite{Nak} deal (for finite systems) with variational calculations of the 2-matrix. Recent HEG papers are \cite{Sun} and \cite{Loos} and refs. within. \\ 

\noindent
Here is roughly sketched, what will be presented in Secs. II, III, IV in more detail.
In this paper it is aimed to present $\chi$ in the RPA with all those terms of $\chi$ contributing to the correlation energy $e_{\rm corr}$ terms 
up to $r_s^2\ln r_s$ and $r_s^2$. The correctness of this $\chi$ is controlled/checked by deriving $n(k)$ and $S(q)$. A comparison is performed between these 'new' RPA results with the 'old' ones of Daniel/Vosko 
\cite{Da}, Kulik \cite{Ku}, and Kimball \cite{Kim3}, respectively. 
For this purpose the diagrams of Figs.1-5 are needed. Fig.1 shows the Yukawa-like screening of the bare (long-range) Coulomb repulsion $v(q)=q_c^2/q^2$, replacing it by the frequency-dependent interaction 
$v(q,\eta)= v(q)/[1+v(q)Q(q,\eta)]$, where 
$Q(q,\eta)$ is the particle-hole propagator (\ref{C3}). The bare Coulomb repulsion shows up in divergencies: for  the  correlation energy $e_{\rm corr}$ see 
\cite{Hei},
for $n(k)$ and $S(q)$ see \cite{Zie3}. These divergencies are eliminated through Macke's partial summation of diagrams. Fig.2a shows the interaction of 2 particle-hole lines running from $1'$ to 1 and from $2'$ to 2. 
This is called "d" = direct diagram to distinguish it from "x" = exchange diagram of Fig.3a with one line
running from $1'$ to 2 and another one from $2'$ to 1. These lowest-order RPA terms of Figs. 2a and 3a are called $\chi_{\rm dr}$ and $\chi_{\rm xr}$, respectively, "r" = ring-diagram summation of Fig. 1. 
The cumulant matrices $\chi_{\rm dr,xr}$ yield the CPDs $h_{\rm dr,xr}$ 
and the CSFs $C_{\rm dr,xr}$ of Figs. 2b and 3b and the interaction energies of Figs. 2c and 3c. But how to obtain $\Delta n(k)$? Using the self-energy $\Sigma(k,\omega)$, the lowest-order $\Sigma$ - diagrams are in 
Figs.4c and 5c: $\Delta n(k)=n_{\rm r}(k)+n_{\rm x}(k)+\cdots$. Of course, RDM-theory must end up with the
same results. Indeed, $n_{\rm r}(k)$ of Fig.4c comes from $\chi_{\rm xr}$ of Fig.3a by the contraction $2'=2$ and $\int d2$, indicated in Fig.4b by a 
small circle with the final result $n_{\rm r}(k)$ of Fig.4c. Similarly, $n_{\rm x}(k)$ of Fig.5c comes from the ladder diagram (in its exchange version) of Fig.5a, contraction makes Fig.5b with the 
final result $n_{\rm x}(k)$ of Fig.5c. Contraction in terms of diagrams mean to transform a 2-line diagram into a 1-line-diagram. This study opens the door for variational 2-matrix calculations and for 
discussing HEG in terms of cumulant geminals and their weights.\\

\noindent
The outline of the paper is as follows. In Sec.II the basic definitions of RDMs, their cumulant decompositions and the contraction SR are presented.
It is derived in detail how the 2-body quantities $S(q)$ and $g(r)$ with its cumulant pedants $C(q)$ and $h(r)$, together with the 1-body quantity $n(k)$,  follow from the cumulant 2-body matrix $\chi$. 
The decisive contraction procedure is in Eqs.(\ref{b25})-(\ref{b28}).
In Secs.III and IV the lowest-order cumulant 2-matrices and what follows from them are presented. Sec.V gives a summary and an outlook. App.A explains the exponential-linked-diagram theorem as the base 
for a systematic cumulant decompositions. App.B deals with SF properties in terms of the coalescing cusp and curvature theorems, of the plasmon SR, and of the inflexion-point trajectory. 
Table 1 and 2 contain the small- and large $q$-behavior of the CSFs and SFs (similar as in \cite{GGSB}). In App.C the particle-hole propagator, the Macke function $I(q)$ and contour integrations are described. 
App.D lists characteristic correlation parameters, vanishing for $r_s\to 0$. \\

\section{Basic definitions and relations}
\setcounter{equation}{0}

\noindent
{\bf 2.1 The 1-matrix and the 2-matrix: definitions and properties} \\
\noindent
The starting point is the definition of the 1-matrix $\gamma_1$ and of the 2-matrix $\gamma_2$ for the HEG-GS in terms of creation and annihilation operators, 
$\psi_1^\dagger$ and $\psi_1$, respectively: 
\begin{equation}\label{b1}
\gamma_1(1|1')=\langle\psi_{1'}^\dagger\psi_1\rangle\ , \quad \gamma_2(1|1',2|2')=\langle\psi_{1'}^\dagger\psi_{2'}^\dagger\psi_2\psi_1\rangle\ .
\end{equation}
The shorthands $1=(\vek r_1, \sigma_1)$ and $\int d1 = \sum\limits_{\sigma_1}\int d^3r_1$ are used. The hermiticity is obvious and Fermi symmetry means $\gamma_2\to -\gamma_2$ 
if 1 and 2 or 1' and 2' are interchanged. ${\hat N} =\int d1 \ \psi_1^\dagger\psi_1$ is the total particle-number operator. The bra vector $"\langle"$ as well as the ket vector "$\rangle$" are
 $N$-body states, therefore $\psi_1\rangle$ and $\langle\psi_1^\dagger$ are $(N-1)$-body states, thus
\begin{equation}\label{b2}
\int d1 \ \langle \psi_1^\dagger\psi_1\rangle=N\ , \quad \int d1 d2 \ \langle \psi_1^\dagger\psi_2^\dagger\psi_2\psi_1\rangle=N(N-1)
\end{equation}
holds for the normalization. The contraction SR
\begin{equation}\label{b3}
\int d2\ \langle \psi_1^\dagger\psi_2^\dagger\psi_2\psi_{1'}\rangle=\gamma_1(1|1')(N-1)
\end{equation}
describes how the 1-matrix $\gamma_1$ for finite $N$ results from the 2-matrix $\gamma_2$. \\

\noindent
It follows the spin structure of $\gamma_1$, the dimensionless 1-matrix $f(r)$ and its FT:
\begin{eqnarray}
\gamma_1(1|1')=\frac{1}{2}\delta_{\sigma_1,\sigma_1'}\gamma_1(\vek r_1|\vek r_1')\ , \quad \gamma_1(\vek r_1|\vek r_1')=\rho f(r_{11'})\ , \quad 
r_{11'}=|\vek r_1-\vek r_1'|\ ,  \quad \\ 
f(r)=\int\limits_0^\infty d(k^3)\ \frac{\sin kr}{kr}\ n(k)\ , \quad f(0)=\int\limits_0^\infty d(k^3)\ n(k)=1\ ,
 \nonumber
\end{eqnarray}
$f'(0)=0$, $f''(0)=-2t/3$ (with $t_0=3/10$ and $f_0''(0)=-1/5$ for $r_s=0$). Its natural orbitals are plane waves, because the system is extended and homogeneous. 
$n(k)$ is the momentum distribution (MD), $\vek k$ and $\vek r$ are measured in units of $k_{\rm F}$ and $k_{\rm F}^{-1}$, respectively. \\

\noindent
The symmetric and antisymmetric 2-body spin functions
\begin{equation}\label{b5}
\delta_{\pm}(\sigma_1|\sigma_1',\sigma_2|\sigma_2')= \frac{1}{2}(\delta_{\sigma_1,\sigma_1'}\delta_{\sigma_2,\sigma_2'}\pm\delta_{\sigma_1,\sigma_2'}
\delta_{\sigma_2,\sigma_1'}) \quad {\rm with} \quad \frac{1}{4}\sum\limits_{\sigma_{1,2}} \delta_\pm(\sigma_1|\sigma_1,\sigma_2|\sigma_2)=\left\{{3/4 \atop 1/4}\right.
\end{equation}
enter the spin structure of $\gamma_2(1|1',2|2')$  as  
\begin{equation}\label{b6}
\gamma_2(1|1',2|2')= \delta_{-}(\sigma_1|\sigma_1',\sigma_2|\sigma_2')\ \gamma_2^+(\vek r_1|\vek r_1',\vek r_2|\vek r_2')
+ \delta_{+}(\sigma_1|\sigma_1',\sigma_2|\sigma_2')\ \gamma_2^-(\vek r_1|\vek r_1',\vek r_2|\vek r_2')
\end{equation}
\cite{Zie12,Zie8}. It defines - together with the basic definition (\ref{b1}) - the singlet 2-matrix $\gamma_2^+$ and the triplet 2-matrix $\gamma_2^-$, being (in space)
symmetric and antisymmetric for the interchanges $\vek r_1\leftrightarrow \vek r_2$ or $\vek r_1'\leftrightarrow \vek r_2'$, respectively. The normalizations
\begin{equation}\label{b7}
\int d^3r_1 d^3r_2\ \gamma_2^\pm(\vek r_1|\vek r_1,\vek r_2|\vek r_2)=\frac{N}{2}\left(\frac{N}{2}\pm 1\right)
\end{equation}
follow from the contractions
\begin{equation}\label{b8}
\int d^3r_2\ \gamma_2^\pm(\vek r_1|\vek r_1',\vek r_2|\vek r_2)=\frac{1}{2}\ \gamma_1(\vek r_1|\vek r_1')\ \left(\frac{N}{2}\pm 1\right)\ ,
\end{equation}
which are in their turn consequences of the total normalization (\ref{b2}) and the singlet-triplet representation (\ref{b6}). 
Eqs.(\ref{b7}) and (\ref{b8}) show the necessity to find an equivalent writing for $\gamma_2$, which allows the thermodynamic limit with $N\to\infty$, $\Omega\to \infty$, $\rho=N/\Omega={\rm const}$. 
This is just done by means of an additive decomposition of $\gamma_2$ in a Hartee-Fock like term $\gamma_{\rm HF}$, which is defined in terms of the (full, exact) 1-matrix $\gamma_1$ only, and 
a non-reducible remainder - the cumulant 2-matrix $\gamma_{\rm c}$. Thus $\gamma_2=\gamma_{\rm HF}-\gamma_{\rm c}$ as the first step of a more general expansion, see App.A. \\

\noindent
{\bf 2.2 The HF approximation} \\
\noindent
The 2-matrix in a Hartree-Fock like approximation (in terms of the exact 1-matrix only) is
\begin{equation}\label{b9}
\gamma_{\rm HF}(1|1',2|2')= \gamma_1(1|1')\gamma_1(2|2')-\gamma_1(1|2')\gamma_1(2|1')\ . 
\end{equation}
This may by also addressed "exact exchange".
Using Eq.(2.4), the same structure (\ref{b6}) as for the total 2-matrix $\gamma_2(1|1',2|2')$ results:
\begin{equation}\label{b10}
\gamma_{\rm HF}(1|1',2|2')=\delta_{-}(\sigma_1|\sigma_1',\sigma_2|\sigma_2')\gamma_{\rm HF}^+(\vek r_1|\vek r_1', \vek r_2|\vek r_2')
\ +\ \delta_{+}(\sigma_1|\sigma_1',\sigma_2|\sigma_2')\gamma_{\rm HF}^-(\vek r_1|\vek r_1', \vek r_2|\vek r_2')
\end{equation}
with
\begin{eqnarray}
\gamma_{\rm HF}^\pm(\vek r_1|\vek r_1', \vek r_2|\vek r_2')&=& \gamma_1(\vek r_1|\vek r_1')\gamma_1(\vek r_2|\vek r_2')\pm
\gamma_1(\vek r_1|\vek r_2')\gamma_1(\vek r_2|\vek r_1')\nonumber \\
&=&\frac{\rho^2}{4}\ [f(r_{11'})f(r_{22'})\pm f(r_{12'})f(r_{21'})]\ .
\end{eqnarray}
Their normalizations
\begin{equation}\label{b12}
\int d^3r_1d^3r_2\ \gamma_{\rm HF}^\pm(\vek r_1|\vek r_1|,\vek r_2|\vek r_2)=\frac{N}{2}\left(\frac{N}{2} \pm 1\right)\mp \frac{N}{2}c
\end{equation}
contain the above introduced L\"owdin parameter $c$, which is originally defined by 
\begin{equation}\label{b13}
1-c=\frac{2}{N}\int d^3r_1d^3r_2\ \frac{1}{4}\ |\gamma_1(\vek r_1|\vek r_2)|^2=\alpha^3\int\limits_0^\infty d(r_{12}^3)\ \frac{1}{2}\ f^2(r_{12})=\int\limits_0^\infty d(k^3)\ n^2(k)\ .
\end{equation}
Note $\rho \int d^3r_1=\rho\Omega=N$ and note that the homogeneous density $\rho$, measured in units of $k_{\rm F}^3$, equals $1/(3\pi^2)$ and that 
$d^3r_{12}/(3\pi^2)=\alpha^3 d(r_{12}^3)$, $\alpha = [4/(9\pi)]^{1/3}$.  \\ 

\noindent
The contractions of $\gamma_{\rm HF}^\pm$ are 
\begin{equation}\label{b14}
\int d^3r_2\ \gamma_{\rm HF}^\pm(\vek r_1|\vek r_1',\vek r_2|\vek r_2)=\frac{1}{2}\gamma_1(\vek r_1|\vek r_1')\left(\frac{N}{2}\pm 1\right) 
\mp\frac{1}{2}\ c(\vek r_1|\vek r_1')\ ,
\end{equation} 
where
\begin{equation}\label{b15}
c(\vek r_1|\vek r_1')=\rho \int\frac{d^3k}{4\pi/3}\ c(k)\ {\rm e}^{{\rm i}\vek k\vek r_{11'}}\ ,\ c(k)=n(k)[1-n(k)]\ , \int d^3r\ c(\vek r|\vek r)=Nc\ .
\end{equation}
$c(k)$ and $c(\vek r,\vek r')$ are referred to as "L\"owdin function" and "L\"owdin matrix", respectively. [$c(k)$  appears in \cite{Her2} as $-2e_k$].  \\

\noindent
{\bf 2.3 The cumulant decomposition and cumulant geminals} \\
\noindent
With this prelude, the above mentioned decomposition
\begin{equation}\label{b16}
\gamma_2(1|1',2|2')= \gamma_{\rm HF}(1|1',2|2')-\gamma_{\rm c}(1|1',2|2')
\end{equation}
defines the cumulant 2-matrix $\gamma_{\rm c}$. Its spin-structure is similar as in (\ref{b6}) and (\ref{b10}) 
\begin{equation}\label{b17}
\gamma_{\rm c}(1|1',2|2')= \delta_{-}(\sigma_1|\sigma_1',\sigma_2|\sigma_2')\ \gamma_{\rm c}^+(\vek r_1|\vek r_1',\vek r_2|\vek r_2')
+ \delta_{+}(\sigma_1|\sigma_1',\sigma_2|\sigma_2')\ \gamma_{\rm c}^-(\vek r_1|\vek r_1',\vek r_2|\vek r_2')
\end{equation}
and it defines (through $\gamma_{\rm 2}^\pm=\gamma_{\rm HF}^\pm-\gamma_{\rm c}^\pm$) the cumulant 2-matrices  $\gamma_{\rm c}^\pm$ with the normalization
\begin{equation}\label{b18}
\int d^3r_1d^3r_2\ \gamma_{\rm c}^\pm(\vek r_1|\vek r_1,\vek r_2|\vek r_2)=\mp\frac{N}{2}c\ ,
\end{equation}
following from the normalizations (\ref{b7}) and (\ref{b12}) . This ensures the normalization of $\gamma_{\rm c}$ to be $\int d1d2\ \gamma_{\rm c}(1|1,2|2)=Nc$, as it should. With the help of Eqs.(\ref{b8}) and (\ref{b14}) 
the contractions of $\gamma_{\rm c}^\pm$ are
\begin{equation}\label{b19}
\int d^3r_2\ \gamma_{\rm c}^\pm(\vek r_1|\vek r_1',\vek r_2|\vek r_2)=\mp\frac{1}{2}\ c(\vek r_1|\vek r_1')\ .
\end{equation}
Eqs.(\ref{b18}) and (\ref{b19}) show that the $\chi_{\rm c}^\pm $ are size-extensively normalized and contracted. 
With $\gamma_{\rm c}^\pm =\D \frac{\rho^2}{4}\chi_\pm$, the dimensionless cumulant 2-matrices $\chi_\pm $ are introduced. Their normalization follows from (\ref{b18}) as 
\begin{equation}\label{b20}
\alpha^3\int\limits_0^\infty d(r_{12}^3)\ \chi _\pm(\vek r_1|\vek r_1,\vek r_2|\vek r_2)=\mp2c\ . 
\end{equation}
The introduction of the singlet/triplet matrices $\gamma_{\rm c}^\pm$ and $\chi_\pm$, is only a necessary intermediate step to present its spin-parallel/antiparallel components $\chi_{\rm p,a}$ in terms  of 
linked diagrams. For diagonal spins $\sigma_{1,2}'=\sigma_{1,2}$ it is $\delta_\pm(\sigma_1|\sigma_1,\sigma_2|\sigma_2)=(1\pm\delta_{\sigma_1,\sigma_2})/2$. So for antiparallel spins  
($\sigma_1=-\sigma_2$, hence $\delta_{\sigma_1,\sigma_2}=0$), respectively  parallel spins ($\sigma_1=\sigma_2$, hence $\delta_{\sigma_1,\sigma_2}=1$) the spin structure (\ref{b17}) leads to 
\begin{equation}\label{b21}
\chi_{\rm a}=\frac{1}{2}[\chi_++\chi_-]\ ,\  \chi_{\rm p}=\chi_-    \quad \leftrightarrow \quad \chi_+=2\chi_{\rm a}-\chi_{\rm p}\ ,\  \chi_-=\chi_{\rm p}\ .   
\end{equation}
The arguments of these matrices are $(\vek r_1|\vek r_1',\vek r_2|\vek r_2')$. Their diagonals $(\vek r_1|\vek r_1, \vek r_2|\vek r_2)$ determine the normalizations 
\begin{equation}\label{b22}
\alpha^3\int\limits_0^\infty d(r_{12}^3)\ \frac{1}{2}\chi_{\rm a}=0, \quad \alpha^3\int\limits_0^\infty d(r_{12}^3)\ \frac{1}{2}\chi_{\rm p}=c  \quad \leftrightarrow \quad 
\alpha^3\int\limits_0^\infty d(r_{12}^3)\ \frac{1}{2}\chi_\pm=\mp c\ .  
\end{equation}
The singlet/triplet matrices $\chi_\pm$ define symmetric and antisymmetric cumulant geminals $\psi_K^\pm (\vek r_1,\vek r_2)$ with corresponding weights $\nu_K^\pm$ and with $\Sigma_K\nu_K^\pm=\mp c$ as total weights.
Which properties do these geminals have? Are they discrete (bound) or/and continuous (scattering) states, the latter with phase shifts? Is there a kind of an aufbau principle for the $\nu_K^\pm$?\\

\noindent
{\bf 2.4 Linked Diagrams}\\
\noindent
As mentioned above, the cumulant 2-matrices $\chi_\pm$ are given by linked diagrams. To each symmetrized direct diagram  
${\rm d}_1\widehat{=}\chi_{{\rm d}_1}(\vek r_1|\vek r_1',\vek r_2|\vek r_2'),\ {\rm d}_2, \cdots$  with 2 open 
particle-hole lines (one running from $\vek r_1'$ to $\vek r_1$ and another one from $\vek r_2'$ to $\vek r_2$) belongs an exchange diagram 
${\rm x}_1\widehat{=}\chi_{\rm x_1}(\vek r_1|\vek r_1',\vek r_2|\vek r_2')=\chi_{\rm d_1}(\vek r_1|\vek r_2',\vek r_2|\vek r_1'),\ {\rm x}_2, \cdots$, such that for 
$\vek r_2'=\vek r_2$ and $\int d^3 r_2$ there is only one line running from $\vek r_1'$ to $\vek r_1$ (like the diagrams of the self-energy $\Sigma$, cf \cite{Zie11}).
With these building elements $\chi_{\rm d}=\chi_{{\rm d}_1}+\chi_{{\rm d}_2}+\cdots $ and $\chi_{\rm x}=\chi_{{\rm x}_1}+\chi_{{\rm x}_2}+\cdots $ the 
singlet/triplet components are $\chi_\pm=\chi_{\rm d}\pm\chi_{\rm x}$ and from (\ref{b21}) follow the decisive relations
\begin{equation}\label{b23}
\chi_{\rm a}=\chi_{\rm d}\ , \quad \chi_{\rm p}=\chi_{\rm d}-\chi_{\rm x}\ \quad \curvearrowright \quad \chi_\pm = \chi_{\rm d}\pm \chi_{\rm x}\ ,
\end{equation}
again with the arguments $(\vek r_1|\vek r_1',\vek r_2|\vek r_2')$. The spin-averaged sum $\chi$ can be written as  
\begin{equation}\label{b24}
\chi=\frac{1}{2}[\chi_{\rm a}+\chi_{\rm p}] \quad {\rm or} \quad \chi=\frac{1}{4}\chi_++\frac{3}{4}\chi_- \quad {\rm or} \quad \chi=\chi_{\rm d}-\frac{1}{2}\chi_{\rm x}\ .  
\end{equation}
The a- and p-components are equally weighted, whereas the
singlet- and triplet-components have the weights 1/4 and 3/4, and the d-and x-components have the "weights" 1 and -1/2, respectively. From Eqs.(\ref{b23}) and (\ref{b24}) follow corresponding relations for the PDs 
$g_{\rm a,p}$ and their cumulant correspondings $h_{\rm a,p}$. Note that for their FTs $S_{\rm a,p}$ and $C_{\rm a,p}$ it holds $S=S_{\rm a}+S_{\rm p}$ and $C=C_{\rm a}+C_{\rm p}$. \\  

\noindent
Now the question is: how real physical properties like the PD's $g_{\rm a,p}(r)$ (by taking the {\it diagonal} elements) as well as the MD $n(k)$ 
(by Fourier transforming the contracted {\it off-diagonal} elements) follow from the (more formal) key quantities $\chi_{\rm a,p}$, respectively $\chi_{\rm d,x}$. \\ 

\noindent
{\bf 2.5 Pair Densities and Structure Factors} \\
\noindent
With the cumulant PDs $ h_{\rm a}(r_{12})=\chi_{\rm a}(\vek r_1|\vek r_1,\vek r_2|\vek r_2)$ and $h_{\rm p}(r_{12})=\chi_{\rm p}(\vek r_1|\vek r_1,\vek r_2|\vek r_2)$ and
their Fourier transforms $C_{\rm a,p}(q)$ following from (\ref{a9}) and (\ref{a10}), the PDs $g_{\rm a,p}(r)$ and SFs $S_{\rm a,p}(q)$ are given by (1.6)-(\ref{a8}). This shows that for the spin-parallel quantities, the functions $f^2(r)$ and its FT $F(q)$ are needed, cf App.B.
 So from the cumulant 2-matrices $\chi_{\rm a,p}(\vek r_1|\vek r_1',\vek r_2|\vek r_2') $ follow not only the spin-antiparallel PD 
$g_{\rm a}(r)$, but also the spin-parallel PD $g_{\rm p}(r)$ (supposed the MD $n(k)$ has been calculated first, cf next paragraph). Rigorous theorems for $C_{\rm a,p}(q\to 0)$, $C_{\rm a,p}(q\to\infty)$, and 
$\int\limits_0^\infty d(q^3)C_{\rm a,p}(q)$ are summarized in App.B. \\
   
\noindent
{\bf 2.6 Contraction Sum Rule and Momentum Distribution}\\
\noindent
In terms of the L\"owdin function $c(k)=n(k)[1-n(k)]$ and of the contracted cumulant matrix (the notation $~~^\smallfrown~~$ means $\vek r_2'=\vek r_2$ and $\int d^3r_2\cdots$)
\begin{equation}\label{b25}
{\rm \textroundcap{$\chi$}}_\pm(\vek r_1|\vek r_1')=\int\frac{d^3r_2}{2\cdot 3\pi^2}\ \chi_\pm(\vek r_1|\vek r_1',\vek r_2|\vek r_2) \quad \curvearrowright \quad
{\rm \textroundcap{$\chi$}}_\pm(k)=\int\frac{d^3r_{11'}}{2\cdot 3\pi^2}\ {\rm e}^{-{\rm i}\vek k\vek r_{11'}}\
{\rm \textroundcap{$\chi$}}_\pm(\vek r_1|\vek r_1')
\end{equation}  
the contraction SR (\ref{b19}) can be written as ${\rm \textroundcap{$\chi$}}_\pm(k)=\mp c(k)$. If the expressions (\ref{b21}) and (\ref{b23}) are contracted and Fourier transformed, then
\begin{eqnarray}\label{b26}
{\rm \textroundcap{$\chi$}}_{\rm d}(k)=\frac{1}{2}[-c(k)+c(k)]=0\ , \quad\quad
{\rm \textroundcap{$\chi$}}_{\rm x}(k)=\frac{1}{2}[-c(k)-c(k)]=-c(k)\ .
\end{eqnarray}
This is the contraction SR, which allows to calculate $n(k)$ from 
\begin{equation}\label{b27}
{\rm \textroundcap{$\chi$}}_{\rm x}(k)=\int\frac{d^3r_{11'}}{2\cdot 3\pi^2}\ {\rm e}^{-{\rm i}\vek k\vek r_{11'}}
\int\frac{d^3r_2}{2\cdot 3\pi^2}\ \chi_{\rm x}(\vek r_1|\vek r_1',\vek r_2|\vek r_2)\ ,
\end{equation}
supposed $\chi_{\rm x}$ is approximately known. The 'direct' diagrams $\textroundcap{$\chi$}_{\rm d}(k)$ with 2 lines do not contribute to $n(k)$,
which alone derives from the 1-line diagram ${\rm \textroundcap{$\chi$}}_{\rm x}(k)$. 
If $n_0(k)=\Theta(1-k)$ means the MD of the ideal Fermi gas, then $n(k)=n_0(k)+\Delta n(k)$ defines its interaction induced correction $\Delta n(k)\gtrless 0$ for $k\gtrless 1$ being normalized as 
$\int d^3k\ \Delta n(k)=0$. Thus the contraction SR (\ref{b26}) reads finally
\begin{equation}\label{b28}
\mp \Delta n(k)+[\Delta n(k)]^2={\rm \textroundcap{$\chi$}}_{\rm x}(k)\ \quad {\rm for} \quad k\gtrless 1\ ,  
\end{equation}
what simplifies as $\Delta n(k\gtrless 1)\approx\mp {\rm \textroundcap{$\chi$}}_{\rm x}(k)$ for small correlation tails $\Delta n(k)$, as this is the case for the RPA with ${\rm \textroundcap{$\chi$}}_{\rm x}(k)={\rm \textroundcap{$\chi$}}_{\rm xr}(k)+\cdots$ and
$\Delta n(k)=\Delta n_{\rm r}(k)+\cdots$. The ellipsis mean beyond-RPA terms.
From a diagrammatical point of view one may expect, that already ${\rm \textroundcap{$\chi$}}_{\rm x}(k)$ is the MD, but this is not the case: as derived above, a factor ${\rm sign}(1-k)$ has to be added, to transform ${\rm \textroundcap{$\chi$}}_{\rm x}(k)$ to $\Delta n(k)$.
So contraction means $\vek r_2'=\vek r_2$, $\int d^3r_2$ {\it and} adding a factor sign$(1-k)$. - From $\Delta n(k)$ follows the L\"owdin parameter $c$, which fixes some CSFs at long wavelengths: $C_{\rm x}(0)=-c$,
$C_{\rm p}(0)=+c$, and $C(0)=c$. \\

\noindent
{\bf 2.7 Energies} \\
\noindent
From $\Delta n(k)$ follows the kinetic correlation energy $\Delta t=\int\limits_0^\infty d(k^3)\Delta n(k)k^2/2$. The same $\Delta n(k)$ makes $\Delta F=F-F_0$ with (\ref{B18}). 
This allows to calculate the interaction energy $v=v_{\rm HF}+v_{\rm c}$, which is built up from the HF part $v_{\rm HF}=v_0+\Delta v_{\rm HF}$ and the cumulant part $v_{\rm c}$:
\begin{equation}\label{b29}
v_0=-\int\limits_0^\infty \frac{d(q^3)}{3\cdot 4}\ \frac{1}{2}F_0(q)v(q)\ ,\
 \Delta v_{\rm HF}=-\int\limits_0^\infty \frac{d(q^3)}{3\cdot 4}\frac{1}{2}\Delta F(q)v(q)\ ,\
v_{\rm c}=-\int\limits_0^\infty \frac{d(q^3)}{3\cdot 4}\ C(q)v(q)\ .
\end{equation} 
$v_0=-\alpha r_s/(4\pi/3)$ is the HF energy in lowest order. $C=C_{\rm a}+C_{\rm p}$ is the FT of $h=[h_{\rm a}+h_{\rm p}]/2$, cf (\ref{a9}) and (\ref{a10}). 
The cumulant interaction energy $v_{\rm c}=v_{\rm a}+v_{\rm p}$ is part of the interaction correlation energy $v_{\rm corr}=\Delta v_{\rm HF}+v_{\rm c}$. 
Thus, the energy-components $t$ and $v$ are available, defining the total energy $e=t+v$ as a function of $r_s$ with $v=r_sde/dr_s$ and $t=e-v$. The correlation energy is $e_{\rm corr}= \Delta t+\Delta v_{\rm HF}+v_{\rm c}$. \\

\noindent
{\bf 2.8 Summary of Sec. II}\\ 
\noindent
If $\chi_\pm(\vek r_1|\vek r'_1, \vek r_2|\vek r'_2)$ is given in terms of diagrams ${\rm d}_1, {\rm d}_2, \cdots$ with 2 not-closed particle-hole lines (e.g. Fig.2a) and corresponding exchange terms 
${\rm x}_1,{\rm x}_2 ,\cdots $ (e.g. Figs.3a, 5a), then: \\
(i)  it follows $n(k)$ from the contraction SR (2.30) and from $n(k)$ it follows $t$ via (\ref{a2}). \\
(ii) $F(q)$ and $f^2(r)$, which are needed for $v_{\rm HF}$ and $g_{\rm p}(r)$, respectively, follow also from $n(k)$, cf App.B. \\
(iii) $h_{\rm a,p}(r)$ and $C_{\rm a,p}(q)$ and the cumulant interaction energy $v_{\rm c}$ follow from $\chi_{\rm a,p}$. \\ 
(iv) Finally, $g_{\rm a}(r)=1-h_{\rm a}(q)$ and $g_{\rm p}(r)= 1-f^2(r)-h_{\rm p}(r)$. \\
The resulting quantities are the GS-energies $t,\ v=v_{\rm HF}+v_{\rm c},\ e=t+v$ as functions of $r_s$ and the MD $n(k)$ as well as the PDs $g_{\rm a,p}(r)$ for the many-body quantum kinematics on the lowest level. \\

\section{The lowest-order cumulant 2-matrix in RPA}
\setcounter{equation}{0}

\noindent
In the following, these basic definitions and relations are applied to the spin-unpolarized HEG in the high-density limit $r_s\to 0$, as an illustrative example. 
All linked-diagram (direct and corresponding exchange) terms are taken into account, which contribute in the energy to terms $\sim r_s^2[\ln r_s+{\rm const}+O(r_s)]$.
These diagrams are shown in Figs.2a, 3a, 4a, 5a. \\        

\noindent
{\bf 3.1 The direct building block $\chi_{\rm d}$ in RPA} \\
\noindent
$\chi_{\rm dr}$ denotes the direct cumulant 2-matrix $\chi_{\rm d}$ in its ring-diagram approximation of Fig.2a: $\chi_{\rm d}=\chi_{\rm dr}+\Delta\chi_{\rm d}$ with an unkown (hopefully small) correction 
$\Delta\chi_{\rm d}$ beyond RPA. In its position representation, this $\chi_{\rm dr}$ is given by
\begin{equation}\label{c1}
\chi_{\rm dr}(\vek r_1|\vek r_1',\vek r_2|\vek r_2')=\frac{9}{2}\int \frac{d^3q}{4\pi}\ \frac{1}{2}\left \{ 
\int\frac{d\eta}{2\pi{\rm i}}\ v(q,\eta){\hat Q}_1(q,\eta){\hat Q_2}(q,-\eta) 
{\rm e} ^{{\rm i} (\vek k_1\vek r_{11'}-\vek k_2 \vek r_{22'}+\vek q\vek r_{12})} 
+{\rm h.c.}\right \} 
\end{equation}
with the short-hands $\vek r_{12}=\vek r_1-\vek r_2$ and $\vek r_{11'}=\vek r_1-\vek r_1'$. The integral operator $\hat Q_{i}(q,\eta)$ generalizes the particle-hole 
propagator $Q(q,\eta)$, cf (\ref{C4}). The diagonal elements $\chi_{\rm dr}(\vek r_1|\vek r_1,\vek r_2|\vek r_2)=h_{\rm dr}(r_{12})$ define a function
\begin{equation}\label{c2}
h_{\rm dr}(r)=\frac{9}{2}\int\frac{d^3q}{4\pi}\ q\ \frac{\sin qr}{qr} \int\limits_0^\infty \frac{du}{\pi}\ R^2(q,u)w(q,u)\ , 
\end{equation}
which contributes with the RPA-beyond term $\Delta h_{\rm d}$ to the CPD, $h_{\rm d}=h_{\rm dr}+\Delta h_{\rm d}$. When going from (\ref{c1}) to (\ref{c2}),
 the contour integration $\eta\to {\rm i}qu$, which replaces simultaneously the frequency integration by a velocity integration, 
leads to the real function $R(q,u)=Q(q,{\rm i}qu)$ for the particle-hole propagator, cf eg \cite{Zie11}, Eq.(\ref{B1}). The effective interaction $w(q,u)=v(q,{\rm i}qu)=v(q)/[1+v(q)R(q,u)]=q_{\rm c}^2/[q^2+q_{\rm c}^2R(q,u)]$
is the RPA screened Coulomb interaction with a Yukawa-like cut-off $q_{\rm c}^2R(q,u)$. The qualitative behavior of $h_{\rm a}=h_{\rm d}$ is in \cite{Zie2}, Fig.3; it starts at $h_{\rm a}(0)=c_1$ and decays, 
passes a zero-intercept and approaches 0 from below, its normalization is zero, see (\ref{B1}). The FT of $h_{\rm dr}(r)$, namely
\begin{equation}\label{c3}
C_{\rm dr}(q)=\int\frac{d^3r}{3\pi^2}\ {\cos\vek q\vek r}\ \frac{1}{2}h_{\rm dr}(r)=\frac{3}{2\pi}\ q \int\limits_0^\infty du\ R^2(q,u)w(q,u)\ ,
\end{equation}
contributes to the CSF $C_{\rm d}=C_{\rm dr}+\Delta C_{\rm d}$ \cite{Kim3} with a beyond-RPA term $\Delta C_{\rm d}$. How does $C_{\rm dr}(q)$ behave for small and large momentum transfers, $q\ll q_c$ 
and $q\to \infty$, respectively, and in between ? \\

\noindent
For small $q\ll q_{\rm c}$, the dynamic RPA-propagator $R(q,u)$ can be approximated by its static truncation $R_0(u)=1-u \arctan (1/u)$ \cite{Kim3} such that $C_{\rm dr}(q\ll q_{\rm c})=L(y)\ q_{\rm c}/2$ 
with $y=q/q_{\rm c}$ and
\begin{equation}\label{c4}
L(y)=\frac{3}{\pi}\int\limits_0^\infty du\ \frac{R_0^2(u)}{y+\D\frac{1}{y}R_0(u)}=\frac{3}{4}y-\frac{\sqrt 3}{2}y^2-\frac{\sqrt 3}{2}y^2\varphi(y), \quad \varphi(y)=-\frac{9}{10}y^2+\frac{3}{5}y^3+\cdots\ .
\end{equation}
(For the properties of $L(y)$ see \cite{Zie3}.) Thus
\begin{equation}\label{c5}
C_{\rm dr}(q\ll q_{\rm c})=z_{\rm F}^2\frac{3q}{8}-\frac{q^2}{4\omega_{\rm pl}}-\frac{q^2}{4\omega_{\rm pl}}\varphi\left(\frac{q}{q_{\rm c}}\right)+\cdots\ ,   
\end{equation}
in agreement with \cite{Iwa,GGSB,Zec}. The ellipsis represents terms originating from the difference $\Delta R(q,u)=R(q,u)-R_0(u)$. The function $C_{\rm dr}(q)$ starts linearly in a small region ($q\ll q_{\rm c}$), which shrinks and 
finally vanishes for $r_s\to 0$.  The beyond-RPA factor $z_{\rm F}^2$ is added by hand with the following argument. As shown in \cite{Zie13}, the small-$q$ behavior of the HF-function $F(q)$ is given in (\ref{B18}). 
It transfers via the SFs to the CSFs and [together with the assumption $\Delta C_{\rm d}(q\ll q_{\rm c})=-z_{\rm F}^2q^3/32+O(q^4)$] makes them behave correctly for small $q$, according to the plasmon SR (\ref{B3}) and (\ref{B13}), 
see also Table 1 and 2. \\

\noindent
At the other end, the large-$q$ asymptotics does not require the ring diagram summation, so perturbation theory holds and the "descreening" replacement $w(q,u)\approx v(q)$ is possible. With the Macke function 
$I(q)=8\pi q \int\limits_0^\infty du\ R^2(q,u)$ (explicitly given in \cite{Zie5}) it is
\begin{equation}\label{c6}
C_{\rm dr}(q\to\infty)=\frac{I(q)}{(4\pi/3)^2}\frac{\omega_{\rm pl}^2}{q^2}+O(\omega_{\rm pl}^4)\ , \quad \frac{I(q)}{(4\pi/3)^2}=\frac{1}{q^2}+\frac{2}{5}\frac{1}{q^4}+O\left(\frac{1}{q^6}\right), \quad \omega_{\rm pl}^2=\frac{q_{\rm c}^2}{3}\ . 
\end{equation}
"In between" $q\to 0$ and $q\to\infty$ and according to (\ref{a9}), the integral $c_{\rm dr}=\int\limits_0^\infty d(q^3)\ C_{\rm dr}(q)$ contributes to the correlation parameter $c_1=c_{\rm dr}+\cdots$, where the ellpipsis represents the 
missing terms beyond RPA. The behavior of $C_{\rm d}=C_{\rm dr}+\Delta C_{\rm d}$ with the RPA-beyond term $\Delta C_{\rm d}$ (causing the correlation parameters $c_{1,2}$) is summarized in line 1 of Table 1 
with unknown coefficients $s_{4,5}$. \\
  
\noindent
{\bf 3.2 The exchange building block $\chi_{\rm x}$ in RPA} \\
\noindent
As stressed by Geldart et al. \cite{Gel1}, for a consistent small-$r_s$ description the leading exchange terms of $n_{\rm x}(k)$, $S_{\rm x}(q)$ are needed. They follow from $\chi_{\rm x}$ with the 
ring-diagram part
$\chi_{\rm xr}$ (Fig. 3a): $\chi_{\rm x}=\chi_{\rm xr}+\Delta\chi_{\rm x}$ with an unkown (hopefully small) correction
$\Delta\chi_{\rm x}$ beyond RPA. That exchange contribution $\chi_{\rm xr}$, which corresponds to the above $\chi_{\rm dr}$, arises from the exchange 
$\vek r_1'\leftrightarrow \vek r_2'$, i.e. from 
$\chi_{\rm xr}(\vek r_1|\vek r_1',\vek r_2|\vek r_2')=\chi_{\rm dr}(\vek r_1|\vek r_2',\vek r_2|\vek r_1')$, thus
\begin{equation}\label{c7} 
\chi_{\rm xr}(\vek r_1|\vek r_1',\vek r_2|\vek r_2')=\frac{9}{2}\int \frac{d^3q}{4\pi}\ \frac{1}{2}\left \{ 
\int\frac{d\eta}{2\pi{\rm i}}\ {\hat Q}_1(q,\eta){\hat Q}_2(q,-\eta)\ v(q,\eta)\
{\rm e}^{{\rm i}(\vek k_1\vek r_{12'}-\vek k_2 \vek r_{21'}+\vek q\vek r_{12})}+{\rm h.c.}\right \}\ . 
\end{equation}
The diagonal elements $h_{\rm xr}(r_{12})=\chi_{\rm xr}(\vek r_1|\vek r_1, \vek r_2|\vek r_2)$ define a function
\begin{equation}\label{c8}
h_{\rm xr}(r)
=\frac{9}{2}\int \frac{d^3q}{4\pi}\ {\rm Re}\int \frac{d\eta}{2\pi{\rm i}}\ \hat Q_1(q,\eta)\hat Q_2(q,-\eta)\ v(q,\eta) 
\frac{\sin kr}{kr}|_{\vek k=\vek k_1+\vek k_2+\vek q}\ , 
\end{equation}
which contributes to the CPD $h_{\rm x}=h_{\rm xr}+\Delta h_{\rm x}$. Its FT  
\begin{equation}
C_{\rm xr}(q)=\int \frac{d^3r}{3\pi^2}\ {\cos \vek q \vek r}\ \frac{1}{2}h_{\rm xr}(r)
= \frac{3}{2}\ {\rm Re} \int \frac{d\eta}{2\pi{\rm i}}\ {\hat Q}_1(k,\eta){\hat Q}_2(k,-\eta)\ v(k,\eta) |_{\vek k=\vek k_1+\vek k_2+\vek q} \nonumber 
\end{equation}
\noindent
contributes to the CSF $C_{\rm x}=C_{\rm xr}+\Delta C_{\rm x}$ with the RPA-beyond term $\Delta C_{\rm x}$. A careful study shows, that $C_{\rm xr}$ can be written as
\begin{eqnarray}\label{c9}
&C_{\rm xr}(q)&=\frac{\omega_{\rm pl}^2}{(4\pi /3)^2}\int d^3k_1d^3k_2\ \int \frac{k\ du}{\pi}\  
\frac{\varepsilon_1\varepsilon_2}{(k^2u^2+\varepsilon_1^2) (k^2u^2+\varepsilon_2^2)} \cdot \frac{1}{k^2+q_{ \rm c}^2R(k,u)}\ ,  \\
& k_{1,2}<1,&\ |\vek k_{1,2}+\vek q|>1,\  \vek k=\vek k_1+\vek k_2 +\vek q ,\ \varepsilon_1=t(\vek  k_2+\vek q)-t(\vek k_1) ,\ \varepsilon_2=t(\vek k_1+\vek q)-t(\vek k_2)\ .\nonumber
\end{eqnarray}
This is the exchange counterpart to the direct CSF (\ref{c3}), unfortunately not known as an explicit function of $q$. If one assumes for the small-$q$ behavior $C_{\rm xr}(q\ll q_c)=O(q^4)$ and $\Delta C_{\rm x}(q\ll q_c)=-c+O(q^4)$, 
then the correct small-$q$ behavior of $C_{\rm p}=C_{\rm d}-C_{\rm x}$ and $S_{\rm p}$ results, namely $C_{\rm p}(0)=c$, see  (\ref{B2}), 
and consequently $S_{\rm p}(0)=0$, see (\ref{a4}). Line 2 of Table 1 contains so far unknown coefficients $s_{4,5}'$. At the other end, the large-$q$ asymptotics of $C_{\rm xr}(q)$ can be extracted from 
perturbation theory, replacing "r" by its lowest order "1". This allows the $u$-integration, by which appears the well-known energy denominator:
\begin{eqnarray}
\int\frac{dx}{\pi}\frac{\varepsilon_1\varepsilon_2}{(x^2+\varepsilon_1^2)(x^2+\varepsilon_2^2)}=\frac{1}{\varepsilon_1+\varepsilon_2}=\frac{1}{\tau_1+\tau_2}=\frac{1}{\vek q(\vek k_1+\vek k_2+\vek q)}\ .
\nonumber
\end{eqnarray}
Thus, it results similar as  $C_{\rm dr}(q\to\infty)$, namely  
\begin{equation}\label{c10}
C_{\rm xr}(q\to \infty)=\frac{I_{\rm x}(q)}{(4\pi/3)^2}\frac{\omega_{\rm pl}^2}{q^2}+O(\omega_{\rm pl}^4),\quad \quad \frac{I_{\rm x}(q\to \infty)}{(4\pi/3)^2}=\frac{1}{q^2}+\frac{x}{q^4}+O\left(\frac{1}{q^6}\right)  
\end{equation}
with an unknown coefficient $x$. For $I_{\rm x}(q)$ in general and possibly $x=8/5$, see App.C. 
The behavior of $C_{\rm x}=C_{\rm xr}+\Delta C_{\rm x}$ is summarized in line 2 of Table 1. An additional correlation parameter $c_2'$ arises from the RPA-beyond term $\Delta C_{\rm x}$.  \\

\noindent
{\bf 3.3 How the cumulant structure factors $C_{\rm a,p}$ follow from $C_{\rm d,x}$} \\
\noindent
For the spin-antiparallel structure factors it holds $C_{\rm a}=C_{\rm d}$. In RPA, $C_{\rm d}(q)$ is given by (\ref{c3}) and its large-$q$ asymptotics is given by (3.6). 
But beyond RPA the electron-electron coalescing cusp theorem 
$h_{\rm a}'(0)=-\alpha r_s [1-h_{\rm a}(0)]$ (App.B1, \cite{Kim1}) makes the first term decorated with an additional factor $1-c_1=1-h_{\rm a}(0)<1$. One may expect the next term is modified in a similar way, thus
\begin{equation}\label{c11}
C_{\rm a}(q\to\infty)= (1-c_1)\frac{\omega_{\rm pl}^2}{q^4}+\left(\frac{2}{5}+c_2\right)\frac{\omega_{\rm pl}^2}{q^6}+\cdots\ , \quad \omega_{\rm pl}^2=\frac{4\alpha r_s}{3\pi} 
\end{equation} 
with unknown correlation parameters $c_{1,2}$ vanishing for $r_s\to 0$.  
Note that the spin-antiparallel CSF $C_{\rm a}(q\to \infty)$ starts with $1/q^4$. \\ 

\noindent
For the spin-parallel structure factors it holds $C_{\rm p}=C_{\rm d}-C_{\rm x}$, $C_{\rm pr}=C_{\rm dr}-C_{\rm xr}$, and $\Delta C_{\rm p}=\Delta C_{\rm d}-\Delta C_{\rm x}$. The large-$q$ asymptotics of $C_{\rm x}(q)$ 
is conjectured as (with another correlation parameter $c_2'$)
\begin{equation}\label{c12}
C_{\rm x}(q\to\infty)= \left(1-c_1\right)\frac{\omega_{\rm pl}^2}{q^4}+\left(x+c_2'\right)\frac{\omega_{\rm pl}^2}{q^6}+\cdots,
\end{equation}
such that the difference $C_{\rm d}-C_{\rm x}=C_{\rm p}$ starts asymptotically according to
\begin{equation}\label{c13}
C_{\rm p}(q\to \infty)=\left(\frac{2}{5}-x+c_2-c_2'\right)\frac{\omega_{\rm pl}^2}{q^6}+\cdots
\end{equation} 
with $1/q^6$ (not with $1/q^4$). Just this conclusion results also from the electron-electron coalescing curvature theorem with $x=8/5$, cf (\ref{B7}).  At the other end, the small-$q$ behavior $C_{\rm p}(q\ll q_c)$ 
follows from the above derived/assumed expressions for $C_{\rm d}(q\ll q_c)$ and $C_{\rm x}(q\ll q_c)$:
\begin{equation}\label{c14}
C_{\rm p}=C_{\rm d}-C_{\rm x}=c+z_{\rm F}^2\left(\frac{3q}{8}-\frac{q^3}{16}\right)-\frac{q^2}{4\omega_{\rm pl}} + \left[s_4''\left(\frac{q}{\omega_{\rm pl}} \right)^4+s_5''\left(\frac{q}{\omega_{\rm pl}} \right)^5+\cdots\right]
\omega_{\rm pl}\ . 
\end{equation}
[For the qualitative behavior of the corresponding CPD $h_{\rm p}$ see \cite{Zie2}, Fig.4; it starts with $h_{\rm p}(r\ll r_c )=(c_3/2)r^2+\cdots$, see (\ref{B2}), (\ref{B5}), (\ref{B8}).]
A consequence for the total CSF $C=C_{\rm a}+C_{\rm p}=2C_{\rm d}-C_{\rm x}$ is
\begin{equation}\label{c15}
C(q\ll q_{\rm c})=c+z_{\rm F}^2\left(\frac{3q}{4}-\frac{q^3}{16}\right)-\frac{q^2}{2\omega_{\rm pl}}+\left[s_4'''\left(\frac{q}{\omega_{\rm pl}} \right)^4+s_5'''\left(\frac{q}{\omega_{\rm pl}} \right)^5+
\cdots\right]\omega_{\rm pl} +\cdots\  .
\end{equation}
This, used in $S(q)=1-\frac{1}{2}F(q)-C(q)=c+z_{\rm F}^2(\frac{3q}{4}-\frac{q^3}{16})-C(q)+\cdots$, cancels the first three terms. Thus emerges the quadratic term  
 in agreement with the plasmon SR $S(q\ll q_{c})=q^2/(2\omega_{\rm pl})+\cdots$.
Consequently an inflexion point appears near the origin. Its trajectory with $r_s\to 0$ is sketched in App.B. Table 2 gives the small-$q$ and large-$q$ behavior of $S_{\rm a,p}$ . For the small-$q$ behavior of 
$S=S_{\rm a}+S_{\rm p}$ the result (\ref{B13}) of the inflexion-point analysis with further correlation parameters $c_{4,5}$ is used, see Table 2. It remains open to check the above assumption on $C_{\rm x}(q\ll q_c)$. \\

\noindent
{\bf 3.4 How $v$ follows from the cumulant structure factor $C=C_{\rm a}+C_{\rm p}=2C_{\rm d}-C_{\rm x}$} \\ 
\noindent
If (\ref{c3}) and (\ref{c9}) are used in (\ref{b29}), then the cumulant interaction energy $v_{\rm c}=v_{\rm d}+v_{\rm x}=v_{\rm dr}+v_{\rm xr}+\cdots$ with ("r" = RPA, ellipsis = beyond RPA)
\begin{equation}\label{c16}
v_{\rm dr}+v_{\rm xr}=-\frac{1}{2}\int\frac{d^3q}{4\pi}\ v(q)\ [\ C_{\rm dr}(q)-\frac{1}{2}C_{\rm xr}(q)]
\end{equation}
consists of a "direct" part $v_{\rm dr}$ and a corresponding "exchange" part $v_{\rm xr}$. The first one,
\begin{equation}\label{c17}
v_{\rm dr}=-\frac{3}{4\pi}\int\frac{d^3q}{4\pi}\ q \int\limits_0^\infty du\ v(q)R^2(q,u)w(q,u)\ ,
\end{equation}
agrees (of course) with what follows from the total energy $e$ after Macke (1950) \cite{Ma} and Gell-Mann/Brueckner (1957) \cite{GB},
\begin{equation}\label{c18}
e_{\rm dr} =-\frac{3}{4\pi}\int\frac{d^3q}{4\pi}\ q \int\limits_0^\infty du \left \{v(q)R(q,u)-\ln[1+v(q)R(q,u)]\right \}\ ,
\end{equation}
by means of the virial theorem $v=r_s de/dr_s$. The "exchange" part $v_{\rm xr}$ can be simplified as
\begin{equation}\label{c19}
v_{\rm xr}= +\frac{1}{4}\int\frac{d^3q}{4\pi}v(q)C_{\rm x1}(q)+\cdots, \quad  C_{\rm x1}(q)=\frac{3}{(4\pi)^2}I_{\rm x}(q)v(q)\ .
\end{equation}
Namely, whereas in $C_{\rm xr}(q)$ the screened interaction $v(q,\eta)$ is necessary to remove the long-range divergencies
of the bare interaction $v(q)$, this is not the case for $v_{\rm x2}$, which has been calculated by Onsager/Mittag/Stephen (1966) \cite{Ons} with the result
\begin{equation}\label{c20}
v_{\rm x2}=\frac{3}{4}\int\frac{d^3q}{(4\pi)^3}\ v^2(q)I_{\rm x}(q)=2\left[\frac{1}{6}\ln 2-\frac{3}{4}\frac{\zeta(3)}{\pi^2}\right](\alpha r_s)^2\ .
\end{equation}
Note that $v_{\rm x2}=2\ e_{\rm x2}$. \\

\noindent
The agreement of (\ref{c18}) with the results of Macke and Gell-Mann/Brueckner \cite{Ma,GB} and of (\ref{c20}) with the calculations of Onsager et al. \cite{Ons} confirms also the above expressions for 
$h_{\rm a,p}(r)$ and $C_{\rm a,p}(q)$ in the RPA approximations "dr" and "xr". \\

\noindent
{\bf 3.5 How $n_{\rm r}$ follows from the contraction of $\chi_{\rm xr}$} \\
\noindent
From $\chi_{\rm xr}$ of (\ref{c7}) follows also a contribution to
\begin{equation}\label{c21}
{\rm \textroundcap{$\chi$}}_{\rm x}(k)={\rm \textroundcap{$\chi$}}_{\rm xr}(k)+\cdots\ , \quad {\rm 
\textroundcap{$\chi$}}_{\rm xr}(k)= \int\frac{d^3r_{11'}}{2\cdot 3\pi^2}{\rm e}^{-{\rm i}\vek k \vek r_{11'}}\ \int\frac{d^3r_2}{2\cdot 3\pi^2}\ \chi_{\rm xr}(\vek r_1|\vek r_1',\vek r_2|\vek r_2)\ . 
\end{equation}
 As explained in App.D, the MD $n(k)$ starts with a 2nd-order term, hence $v(q,\eta)$ can be replaced by $-v(q)Q(q,\eta)v(q,\eta)$. Therefore
\begin{eqnarray}\label{c22}
\chi_{\rm xr}(\vek r_1|\vek r_1',\vek r_2|\vek r_2)=\frac{9}{2}\int \frac{d^3q}{4\pi}\ \frac{1}{2}\left\{ \int\frac{d\eta}{2\pi{\rm i}}\ [-v(q)Q(q,\eta)v(q,\eta)]\ \cdot \right.  \\
\left. \cdot\ \hat Q_1(q,\eta)\hat Q_2(q,-\eta) {\rm e} ^{-{\rm i}(\vek k_1+\vek k_2+\vek q)\vek r_2}\ {\rm e}^{{\rm i}[(\vek k_1+\vek q)\vek r_1+\vek k_2\vek r_1']}+{\rm h.c.} \right\}\ .  \nonumber 
\end{eqnarray}
The $\vek r_2$-integration and afterwards the $\vek r_{11'}$-integration yield (including prefactors)
\begin{equation}\label{c23}
\left(\frac{(2\pi)^3}{2\cdot 3\pi^2}\right)^2\ \delta (\vek k_1+\vek k_2+\vek q)\ \delta (\vek k+\vek k_2)=\left(\frac{4\pi}{3}\right)^2\delta (\vek k_1+\vek q-\vek k)\delta (\vek k_2+\vek k)\ ,
\end{equation}
so that
\begin{equation}\label{c24}
\textroundcap{$\chi$}_{\rm xr}(k)= -\frac{1}{2}\int \frac{d^3q}{4\pi}\ \frac{1}{2}\left\{ \int\frac{d\eta}{2\pi{\rm i}}\ v(q)Q(q,\eta)v(q,\eta) X(q,\eta) + {\rm c.c.} \right\}
\end{equation}
with
\begin{equation}\label{c25}
 X(q,\eta)=\ 4\pi\ \hat Q_1(q,\eta) \delta(\vek k_1+\vek q-\vek k)\ \cdot  4\pi\ {\hat Q}_2(q,-\eta)\delta(\vek k_2+\vek k)=\frac{\tilde\Theta(\vek k,\vek q)}{[\tau(\vek k,\vek q)-\eta]^2}\ ,
\end{equation}
where $\tilde\Theta(\vek k,\vek q)=1$ for $|\vek k|\gtrless 1, |\vek k+\vek q|\lessgtr 1$ and $0$ otherwise. The last step is derived in App.C4.
This used in (\ref{c24}) gives for the MD-correlation part the result $\Delta n(k)=n_{\rm r}(k)+\cdots$ with 
\begin{equation}\label{c26}
n_{\rm r}(k\gtrless 1)=\pm\frac{1}{2}\int \frac{d^3q}{4\pi}\ \frac{1}{2}\left \{ \int\frac{d\eta}{2\pi{\rm i}}v(q)Q(q,\eta)v(q,\eta)
\frac{\tilde\Theta(\vek k,\vek q)}{[\tau(\vek k,\vek q)-\eta]^2}+{\rm c.c.}\right \}\ . 
\end{equation}
With $\Theta(\vek k,\vek q)=\pm \tilde\Theta(\vek k,\vek q)$ for $|\vek k|\gtrless 1$ and again with $\eta\to {\rm i}q u$, it turns out
\begin{equation}\label{c27}
n_{\rm r}(k)=\frac{1}{2}\int\limits_0^\infty d(q^2)\int\limits_0^\infty \frac{du}{2\pi}\ v(q)R(q,u)w(q,u)\ {\rm Re}\int\limits_{-1}^{+1}\frac{d\zeta}{2}
\frac{\Theta(\vek k,\vek q)}{[(k\zeta+\frac{q}{2})-{\rm i}u]^2}\ ,
\end{equation}
being $\gtrless 0$ for $k\gtrless 1$.
This is exactly the RPA-MD of Daniel/Vosko (1960) and Kulik (1961), as it has been shown in detail in \cite{Zie11} on the way therein from Eqs.(3.34) to (3.38). Also therein at (3.10) it is shown, 
what $n_{\rm r}(k)$ contributes to the kinetic energy $t$ in agreement with the virial theorem (\ref{a1}) and the Macke/Gell-Mann/Brueckner energy (\ref{c18}). The asymptotic behavior is 
$n(k\to \infty)=\omega_{\rm pl}^4/(2k^8)+\cdots$ for RPA and $n(k\to \infty)=(1-c_1)\omega_{\rm pl}^4/(2k^8)+\cdots$ beyond it. This 2nd-order result is reduced by a factor of 1/2 through a still missing 
ladder diagram in its exchange version, as treated in the next section.

\section{The 2nd-order 2-matrix in RPA}
\setcounter{equation}{0}

\noindent
This ladder diagram has a particle-hole line running from $1'$  to 2 and a 2nd line from $2'$ to 1, cf Fig.5a.  
Its contraction $\vek r_2'=\vek r_2$ and $\int d^3r_2$ yields the diagram of Fig.5b, which differs from Fig.5c only by a factor sign$(1-k)$ according to (\ref{b28}). In detail the diagram $\chi_{\rm lxr}$ of Fig.5a is given by 
(\ref{C12}). If for simplicity the RPA screening is neglected (corresponding to the replacement r$\to$2) and a series of laborious contour integrations are carefully performed together with the contraction steps analog (\ref{c21}), it finally yields 
\begin{equation}\label{d1}
{\rm \textroundcap{$\chi$}}_{\rm lx2}(k)=\frac{1}{4}\int\frac{d^3q_1d^3q_2}{(4\pi/3)^2}\frac{\omega_{\rm pl}^4}{q_1^2q_2^2}\frac{\Theta(\vek k, \vek q_{1,2})}{(\vek q_1\cdot\vek q_2)^2}>0 
\quad \curvearrowright\quad n_{\rm x}(k\gtrless 1)= \mp{\rm \textroundcap{$\chi$}}_{\rm lx2}(k)\ .
\end{equation} 
The mentioned factor sign$(1-k)$ transforms ${\rm \textroundcap{$\chi$}}_{\rm lx2}(k)$ to the corresponding MD contribution $n_{\rm x}(k)$, which thus reduces both the particle branch ($k>1$) and the hole 
branch ($k<1$) of the RPA-MD $n_{\rm r}(k)$, Eq.(\ref{c27}). In particular the asymptotics for $k\to \infty$ changes from 1/2 to 1/4 of $\omega_{\rm pl}^4/k^8$ as already mentioned at the end of Sec.III. 
Furthermore it is easy to show by means of the substitutions $\vek k\to -(\vek k+\vek q_1)$ and $\vek q_2\to -\vek q_2$, (i) the zero normalization $\int d^3k\ n_{\rm lx2}(k)=0$  and (ii) the kinetic energy 
\begin{equation}\label{d2}
t_{\rm lx2}=\frac{1}{8}\int\frac{d^3kd^3q_1d^3q_2}{(4\pi/3)^3}\frac{\omega_{\rm pl}^4}{q_1^2q_2^2}\frac{\Theta^-(\vek k,\vek q_{1,2})}{\vek q_1\cdot \vek q_2},
\end{equation} 
where also the identity (\ref{C20}) is used. The result (\ref{d2}) is just the exchange integral, which has been solved analytically by Onsager et al. \cite{Ons}. Whereas (\ref{d2}) is a non-divergent 2nd-order term, 
for $n(k\approx 1)$ an indeed complicated ring-diagram summation is needed, to ameliorate diverging terms. \\

\section{Summary and  outlook}
\setcounter{equation}{0}
\noindent
This is a contribution to the theory of extended many-electron systems in terms of reduced-density matrices (RDMs), which replace $\Phi(1,\cdots,N)$, the many-electron wave functions of finite systems and 
form an infinite hierarchy, 
such that a lower-level RDM follows from the next-higher-level RDM by means of a certain contraction procedure, e.g. from the 2-body matrix (2-matrix) $\gamma_2$ follows the 1-matrix $\gamma_1$. This paper is driven 
by the belief that the {\bf cumulant 2-matrix} $\gamma_{\rm c}$ (dimensionless: $\chi$) has a fundamental meaning. This shows up for extended systems (i) in its size-extensive normalization and contraction and 
(ii) in its (closely related) 
{\it linked}-diagram representation. Besides it is a decisive key quantity, because from it follows both the pair density (PD) and the 1-matrix. Here it is aimed to unveil its secrets for the high-density 
spin-unpolarized homogeneous electron  gas (HEG).  Although this model is only a marginal corner in the complex field of electron correlation, 
one should get generally deeper insight into the many-body correlation by exploiting the concept of cumulant decomposition of RDMs. 
The ground state (GS) energy of an electronic system is a simple functional of $\rho(\vek r, \sigma)$ = electron density, of $n(\vek k, s)$ = momentum distribution (MD), and of 
$g(1,2)$ = pair density (PD) with $1=(\vek r_1,\sigma_1)$. These 1- and 2-body quantities have their common origin in the 2-matrix $\gamma_2(1|1',2|2')$. Whereas the PD simply 
follows from $\gamma_2$ by taking the diagonal elements, the 1-body quantities need a procedure called contraction which means $2'=2$ and $\int d2$. But this procedure leads in the case of extended systems to 
difficulties with the thermodynamic limit (TDL). Namely non-size-extensive expressions 
$\sim N^2$ appear, resulting from the numbers of pairs $N(N-1)$. This is avoided 
by the cumulant decomposition of the 2-matrix $\gamma_2$. This means that $\gamma_2$ is additively decomposed into a HF-like (or exact exchange) term $\gamma_{\rm HF}$, which reduces to a sum of products of the 
(correlated) 1-matrix, 
and a non-reducible remainder, called cumulant matrix $\gamma_{\rm c}$. Its advantage: it is {\it size-extensively} normalizable and contractable. Thus the problems with the TDL are eliminated. This is worked out in 
detail for the HEG with the homogeneous electron density $\rho$, with the MD $n(k)$ and with $S(q)$, the structure factor (SF) [or equivalently the 1-matrix $f(|\vek r_1-\vek r_1'|)$ and the PD $g(r_{12})$, 
respectively]. The 2-body quantities 
$S(q)$ and $g(r)$ are decomposed into HF terms and remainders $C(q)$ and $h(r)$, respectively called cumulant SF (CSF) and cumulant PD (CPD). In detail, this means $g(r)=1-\frac{1}{2}f^2(r)-h(r)$ and 
$S(q)=1-\frac{1}{2}F(q)-C(q)$ with $F(q)$= Fourier transform of $f^2(r)$. The contraction procedure is developed in Eqs.(\ref{b25})-(\ref{b28}): A quadratic equation for $\Delta n(k)=n(k)-n_0(k)$ yields 
the 2 branches for particles ($k>1$) and holes ($k<1$). \\

\noindent
Closely related to the described {\it size-extensivity} of the cumulant 2-matrix is its form (within many-body perturbation theory) in terms of {\it linked} diagrams. Under the impression that 
the triangle of "cumulants $\leftrightarrow$ size-extensivity $\leftrightarrow$ linked diagrams" is a fundamental relation (with an exponential-linked-diagram theorem as a more general mathematical 
background \cite{Zie14,Zie15}), one may curiously ask for the {\bf cumulant geminals} $\psi_K^\pm(\vek r_1,\vek r_2)$ and their weights $\nu_K^\pm$, which diagonalize 
$\gamma_{\rm c}$ \cite{Zie16}, analogous to Eq.(\ref{b16}) in \cite{Zie2}. What are their properties? Is there perhaps a 2-body scheme, which allows to calculate approximately 
these $\psi_K^\pm$ and $\nu_K^\pm$? How they differ from the Overhauser geminals \cite{Over}, which parametrize the PD $g(r)$? One access to this subject are the contracted Schr\"odinger equations 
\cite{Zie15}. But instead of this, in this paper the cumulant 2-matrix of the high-density HEG is derived. The results for $\gamma_{\rm c}$ (dimensionless $\chi$) prove to be correct, because they 
yield the known RPA expressions for $S(q)$, $g(r)$, and $n(k)$. This is the 1st step for further curiousity studies along the line of Overhauser \cite{Over}, 
namely to find the cumulant geminals of the high-density HEG, to be specific: the expressions (\ref{c1}), (\ref{c7}), and (\ref{C12}) have to be diagonalized.   \\   

\noindent
Furthermore, the coalescing cusp and curvature theorems, the plasmon sum rule (and its consequence, the inflexion-point trajectory), as well as the small- and large-$q$ behavior of the CSFs for 
electron pairs with parallel and antiparallel spins are systematically summarized (analog to \cite{GGSB}), see Table 1 and 2. Within this, also the so far assumed asymptotic behavior of the 
structure factors for small and large $q$ is corrected.

\section*{Acknowledgments}
\noindent
The author is grateful to P. Gori-Giorgi, Jian Wang, K. Morawetz, J. Bodyfelt for discussions and hints and acknowledges P. Fulde and the Max Planck Institute for the 
Physics of Complex Systems Dresden for supporting this work and thanks Th. M\"uller and R. Schuppe for technical help. \\  

\begin{appendix}
\section*{Appendix A: Generating functionals and linked diagrams}
\setcounter{equation}{0}
\renewcommand{\theequation}{A.\arabic{equation}}
\noindent
The most compact quantity containing all the secrets of the HEG GS $\rangle$ is the generating functional
\begin{equation}\label{A1}
\gamma[\eta^\dagger|\eta]=\langle \hat P\ {\rm e}^{\int dx[\eta^\dagger(x)\psi(x)+\eta(x)\psi^\dagger(x)]}\rangle
\end{equation}
with $x=(\vek r,\sigma)$ and $\int dx=\sum\limits_{\sigma}\int d^3r$. $\eta(x)$ and $\eta^\dagger(x)$ are Grassmann variables and $\hat P$ makes the normal ordering, i.e. 
orders the creation operators
$\psi^\dagger$ including sign changes to the left. The Volterra coefficients of
$\gamma[\eta^\dagger|\eta]$ are the RDMs $\gamma_1$, $\gamma_2$, ... . The ansatz
$\gamma[\eta^\dagger|\eta]={\rm e}^{\chi[\eta^\dagger|\eta]}$
defines another generating functional, $\chi[\eta^\dagger|\eta]$, the Volterra coefficients of which are the so-called cumulant matrices $\chi_1$, $\chi_2$, ... , such that a hierarchy results with higher level
RDMs expressed by sums of products of lower level RDMs and non-reducible remainders. Thus it holds e.g. $\gamma_1(x|x')=\chi_1(x|x')$, $\gamma_2(x_1|x_1',x_2|x_2')= 
\tilde A \chi_1(x_1|x_1')\chi_1(x_2|x_2')+\chi_2(x_1|x_1',x_2|x_2')$, ... with $\tilde A$ being an antisymmetrizer \cite{Zie14}. Within time-dependent perturbation theory ($S$-matrix 
theory) the cumulant matrices $\chi_1$, $\chi_2$, ...  are given by non-vacuum {\it linked} diagrams only and closely related to this they are size-extensively normalized
i.e. $\int dx_1dx_2\cdots \chi(x_1|x_1,x_2|x_2,\cdots)\sim N$, what allows for extended systems the thermodynamic limit with $N\to \infty$, $\Omega\to\infty$, $\rho=N/\Omega={\rm const}$ \cite{Zie15}. \\

\section*{Appendix B: Structure Factors}
\setcounter{equation}{0}
\renewcommand{\theequation}{B.\arabic{equation}}
\noindent
Here are summarized the normalizations of $C_{\rm a,p}(q)$, $C(q)$ etc. as well as their behavior for small and large arguments. \\

\noindent
{\bf B1: Perfect Screening Sum Rule. Cusp and Curvature Theorems} \\
$S(q)$ starts with $S(0)=0$ and approaches $1$ for $q\to \infty$. $g(r)$ starts with $g(0)\leq 1/2$ and oscillatory approaches $1$ for $r\to \infty$. Note, that 
$g_{\rm a,p}(r)\geq 0 \curvearrowright  g(r)\geq 0$, since they are probabilities. 
The normalizations of $1-S(q)$ and $1-g(r)$ are contained in (\ref{a5}) with $g(0)=(1-c_1)/2$ (defining the correlation parameter $c_1$) and with $S(0)=0$ (the perfect 
screening SR), respectively:
\begin{equation}  
\int\limits_0^\infty d(q^3)\ [1-S(q)]=1+c_1\ , \quad \quad \alpha^3 \int\limits_0^\infty d(r^3)\ [1-g(r)]=1\ . \nonumber
\end{equation}
Similarly, the normalizations of $1-g_{\rm a,p}(r)$, $S_{\rm a}(q)$, and $1-S_{\rm p}(q)$ are in (1.4) fixed with $g_{\rm a}(0)=c_1$, $g_{\rm p}(0)=0$, and 
$S_{\rm a,p}(0)=0$. In terms of cumulant and "a,p" components, this reads as 
\begin{equation}\label{B1}
h_{\rm a}(0)=c_1 \leftrightarrow \int\limits_0^\infty d(q^3)\ C_{\rm a}(q)=c_1\ ,\quad \quad
C_{\rm a}(0)=0   \leftrightarrow \alpha^3\int\limits_0^\infty d(r^3)\frac{1}{2}h_{\rm a}(r)=0\ ,  
\end{equation}
\begin{equation}\label{B2}
h_{\rm p}(0)=0   \leftrightarrow \int\limits_0^\infty d(q^3)\ C_{\rm p}(q)=0\ , \quad \quad
C_{\rm p}(0)=c   \leftrightarrow \alpha^3\int\limits_0^\infty d(r^3)\frac{1}{2}h_{\rm p}(r)=c\ ,
\end{equation}
which follows from the FTs (\ref{a9}) and (\ref{a10}) for $r=0$, respectively for $q=0$.  
With $\frac{1}{2}F(0)=1-c$, the perfect screening SR $S(0)=0$ is hidden in $C(0)=C_{\rm a}(0)+C_{\rm p}(0)=c$. More general is the plasmon SR \cite{Pin,Thom}
\begin{equation}\label{B3}
 S(q\ll q_{\rm c})=\frac{q^2}{2\omega_{\rm pl}}+O(q^4)\ ,
\end{equation}
see Table 2, last line. \\

\noindent
According to the electron-electron coalescing cusp theorem $g_{\rm a}'(0)=\alpha r_s\ g_{\rm a}(0)$ for the spin-antiparallel CPD \cite{Kim1}, it holds
$h_{\rm a}'(0)=-\alpha r_s (1-c_1)$ with 
\begin{equation}\label{B4}
c_1=h_{\rm a}(0)=\frac{2}{5 \pi}\left(\pi^2-3+6\ln 2\right)\ \alpha r_s+2\ \left(3-\frac{\pi^2}{4}\right)\left(\frac{3\alpha}{2\pi}\right)^2\ r_s^2\ln r_s+O(r_s^2)
\end{equation}
\cite{Gel2,Kim3}. Exactly this high-density behavior of $h(0)$ and $g(0)$ results also from the ladder theory as the best method to treat short-range correlation \cite{Qi,Cio1,Ya}.
By means of the Kimball trick \cite{Kim1} it follows the 1st term of Eq.(\ref{c11}),
whereas the 2nd term follows from $C_{\rm a}(q)=C_{\rm d}(q)\approx C_{\rm dr}(q)$ and from (\ref{c6}). $c_2$ is another correlation parameter vanishing for $r_s\to 0$.  $c_1$ appears also in the large-$k$ asymptotics
$n(k\to \infty)= (1-c_1)\omega_{\rm pl}^4/(4k^8)+\cdots$. Thus short-range correlation determine the large-wave number asymptotics of $n(k)$ \cite{Kim2,YaKa} and 
of $C(q)$ \cite{Kim2,Kim1,Ya}. \\  

\noindent
For the spin-parallel quantities, the Pauli principle makes $g_{\rm p}(0)=g_{\rm p}'(0)=0$ or equivalently $h_{\rm p}(0)= h_{\rm p}'(0)=0$.
Furthermore the electron-electron coalescing curvature theorem says $g_{\rm p}'''(0)=\frac{3}{2}\alpha r_s\  g_{\rm p}''(0)$ or equivalently $h_{\rm p}'''(0)=\frac{3}{2}\alpha r_s\ [2f''(0)+h_{\rm p}''(0)]$. 
With $f''(0)=-\frac{2}{3}t$ and with \cite{Kim5}
\begin{equation}\label{B5}
c_3=h_{\rm p}''(0)=\frac{1}{35 \pi}\ (4\pi^2-5+20\ln 2)\ \alpha r_s+O(r_s^2\ln r_s)\ , 
\end{equation}
it holds $h_{\rm p}'''(0)=\alpha r_s (\frac{3}{2}c_3-2t)$. This determines the large-$q$ asymptotics of $C_{\rm p}(q)$ as shown in the following by means of the Kimball trick \cite{Kim1}. 
First a function $c_{\rm p}(q)\sim 1/q^8$ is defined through $C_{\rm p}(q)=a/(1+q^2)^3+c_{\rm p}(q)$. Then from the FT (\ref{a10}) it follows
\begin{equation}\label{B6}
h_{\rm p}(r)=a\frac{3\pi}{16}(1+r){\rm e}^{-r}+\int\limits_0^\infty d(q^3)\ c_{\rm p}(q)\frac{\sin qr}{qr}\quad \curvearrowright \quad h_{\rm p}'''(0)=a\frac{3\pi}{8}\ .
\end{equation}
With the above curvature theorem it finally results
\begin{equation}\label{B7} 
C_{\rm p}(q\to \infty)=\left(3\ c_3-4t\right)\frac{\omega_{\rm pl}^2}{q^6}+\cdots\ .
\end{equation}
Comparing this with (\ref{c13}), the SR $\frac{2}{5}-x+c_2-c_2'=3c_3-4t$ seems to hold. For $r_s\to 0$ it takes $t\to 3/10$, hence $x=8/5$ [for plausible arguments see text after (\ref{C8})]. \\

\noindent
Because $C_{\rm p}(q\to \infty)$ decays so strong, from (\ref{B6}) it follows in addition to (B.3) the relation
\begin{equation}\label{B8}
c_3=h_{\rm p}''(0)=-\int\limits_0^\infty dq\ q^4\ C_{\rm p}(q)\ . 
\end{equation}
Since $h_{\rm p}''(0)>0$, cf \cite{Kim4,Kim5}, the integral has to be negative. This is in agreement with (\ref{B3}) and (\ref{B7}): a factor $q^2$ in front of $C_{\rm p}(q)$
makes the integral vanishing, cf (\ref{B3}). An additional factor $q^2$ enhances the negative 
asymptotic branch of $C_{\rm p}(q\to \infty)$, cf (\ref{B7}). Besides, from $g_{\rm p}''(0)>0$ it follows $h_{\rm p}''(0)< 4t/3$. 
For $t(r_s)$ cf eg \cite{GGZ}.  \\  

\noindent
{\bf B2: Plasmon SR and inflexion-point trajectory}\\
\noindent
In (\ref{c5}) it has been shown how perturbation theory and RPA makes the plasmon-term $\sim q^2$ arise from the linear term $\sim q$ of the non-interacting system.   
Here it is asked on the contrary, if the interaction is switched off ($r_s\to 0$), how the quadratic behavior $\sim q^2$ is transformed to the linear behavior $\sim q$.
The answer: if $S(q)$ starts for small $q$ according to the plasmon SR \cite{Thom} quadratically with $q^2$, then there must be - already from a topological point of view - an
inflexion point $q_0$, $S(q_0)$, which moves with $r_s\to 0$ towards the origin. So also the linear behavior at the inflexion point is transported to the origin, realizing 
thus the correct behavior of the unperturbed ideal Fermi gas, as it should. Now this qualitative discussion is quantized by the ansatz 
\begin{equation}\label{B9}
S(q\ll q_c)=\frac{q^2}{2\omega_{\rm pl}}[1-a z^2+b z^3+\cdots]\ , \quad z=\frac{q}{\omega_{\rm pl}} 
\end{equation}
and adjusting the $r_s$-dependent coefficients $a$, $b$ appropriately. Namely the power expansion of $S(q)$ around $q\approx q_0$
\begin{equation}\label{B10}
S(q_0+x)= S(q_0)+\frac{1}{1!}S'(q_0)x+\frac{1}{2!}S''(q_0)x^2+\frac{1}{3!}S'''(q_0)x^3+\cdots
\end{equation}
is compared with its "$r_s\to 0$" limit $S_0(x)$. This gives 3 equations, which allows determining the 3 quantities $q_0$, $a$, $b$:
\begin{equation}\label{B11}
\frac{1}{1!}S'(q_0)\to \frac{3}{4}\ , \quad \frac{1}{2!}S''(q_0)=0 \ , \quad \frac{1}{3!}S'''(q_0)\to-\frac{1}{16}
\end{equation}
with the result (for $r_s\to 0$)
\begin{equation}\label{B12}
\frac{q_0}{\omega_{\rm pl}}\to\frac{3}{2}\ , \quad a\to\frac{1}{3^2}\ , \quad b\to\frac{2^2}{3^3 \cdot 5} \quad .
\end{equation}
So (\ref{B9}) takes for $r_s\to 0$ the form (with unknown correlation parameters $c_{4,5}$)\cite{Iwa}
\begin{equation}\label{B13}
S(q\ll q_{\rm c})=\frac{q^2}{2\omega_{\rm pl}}\left[1-\left(\frac{1}{3^2}+c_4\right)\left(\frac{q}{\omega_{\rm pl}}\right)^2+
\left(\frac{2^2}{3^3\cdot 5}+c_5\right)\left(\frac{q}{\omega_{\rm pl}}\right)^3+\cdots\right]+\cdots\ .
\end{equation}
Thus the inflexion point moves towards the origin according to $q_0\to 1.5\ \omega_{\rm pl}$ and 
$S(q_0)\to 0.96\ \omega_{\rm pl}$ with a slope of $0.64$ being smaller than $S_0'(0)=0.75$. Comparison of (\ref{B13}) with (\ref{c15}) shows $s_4'''=\frac{1}{2}(\frac{1}{3^2}+c_4)$ and 
$s_5'''=-\frac{1}{2}(\frac{2^2}{3^3\cdot 5}+c_5)$. - As quoted in \cite{Zie13}, other consequences of (\ref{B9}) are $g(r\gg r_{\rm c})=1-2^5/(\omega_{\rm pl}^4\ r^8)+\cdots$ and 
\begin{equation}\label{B14}
\alpha^3\int\limits_0^\infty dr\ r^4[1-g(r)]=\frac{1}{\omega_{\rm pl}}\ ,\quad 
\alpha^3\int\limits_0^\infty dr\ r^6[1-g(r)]=\frac{2^3 \cdot 5}{3^2}\frac{1}{\omega_{\rm pl}^3}\ . 
\end{equation}
 The equation with the weight $r^4$ is widely used in \cite{Zec}. Remind $1-g(r)=\frac{1}{2}f^2(r)+h(r)$, what relates $\int d^3k\ (dn(k)/dk)^\nu$ to $\int d^3r\ r^\nu h(r)$ for $\nu=2$ and $\nu=4$.\\

\noindent
{\bf B3: The HF-function $F(q)$}\\
\noindent
For the spin-parallel quantities $g_{\rm p}(r)\leftrightarrow S_{\rm p}(q)$ and $h_{\rm p}(r)\leftrightarrow C_{\rm p}(q)$ the functions $f^2(r)$ and its FT $F(q)$ are needed, cf (\ref{a7}), (\ref{a8}) and Table 2:
\begin{equation}\label{B15}
 F(q)=\alpha^3\int\limits_0^\infty d(r^3)\ \frac{\sin qr}{qr}f^2(r)\ , \quad \frac{1}{2}F(0)=1-c\ , \quad \int\limits_0^\infty d(q^3)\ \frac{1}{2}F(q)=1\ .
\end{equation}
$F(q)$ transports the MD $n(k)$, as seen from 
\begin{equation}\label{B16}
 \frac{1}{2}F(q)=\frac{3}{4\pi}\int d^3k\ n(k)n(|\vek k-\vek q|)\ .     
\end{equation}
Thus it is related to the probability of finding a pair of electrons with given relative momentum $\vek q$ \cite{GGP}.
For $r_s=0$ with $n_0(k)=\Theta(1-k)$ the integral takes a simple geometric meaning: it is just the volume of two spherical calottes with the height $h=1-q/2$, thus $\int d^3k \cdots=2\cdot \frac{\pi}{3}h^2(3-h)$ or   
\begin{equation}\label{B17}
\frac{1}{2}F_0(q)=\left[1-\frac{3}{2}\frac{q}{2}+\frac{1}{2}\left(\frac{q}{2}\right)^3\right] \Theta(2-q)\ , \quad \int\limits_0^\infty d(q^3)\ \frac{1}{2}F_0(q)=1\ .
\end{equation}
[By the way, (i) $F_0(q)$ appears in the Overhauser theory as the weight of its PD geminals \cite{Over} and (ii) the singularities of $F(q)$ at $q\to 0$ and $q\to 2$ are studied in \cite{Zie13}.] 
The difference $\Delta F(q)=F(q)-F_0(q)$ results from $\Delta n(k)=n(k)-n_0(k)$. It holds 
\begin{equation}\label{B18}
\frac{1}{2}\Delta F(q\ll q_c)=-c-z_{\rm F}^2\left(\frac{3q}{4}-\frac{q^3}{16}\right)+\ \cdots,\quad \frac{1}{2}\Delta F(q\to \infty)=\frac{1-c_1}{8}\frac{\omega_{\rm pl}^4}{q^8}+\cdots\ ,  
\end{equation}
and $\int\limits_0^\infty d(q^3)\Delta F(q)=0$. \\

\noindent
{\bf B4: The jump discontinuity of $S''(q)$ and $C''(q)$ at $q=2$}\\
\noindent
For transition momenta $|\vek q|=2$, when passing this value from below, the topology changes from 2 overlapping to 2 non-overlapping Fermi spheres. The consequence: 
$\Delta I''(2)=2\pi^2$, cf (C.2) in \cite{Zie5} $\curvearrowright \Delta C''(2)=(3\omega_{\rm pl}/4)^2+\cdots$ \cite{Zie3}. The ellipsis represents exchange and beyond-RPA terms. For $S''(2)$ the relation 
$\frac{1}{2}\Delta F''(2)=-3z_{\rm F}^2/4$, cf (5.2) in \cite{Zie13}, 
has to be taken into account. Thus in lowest order $\Delta S''(2)=3z_{\rm F}^2/4-(3\omega_{\rm pl}/4)^2+\cdots$, what causes (in the direct space) the Friedel oscillations of $g(r)$. 
It remains open to study the exchange term $\Delta I_{\rm x}''(2)$, which presumably reduces the direct term $\Delta I''(2)$.   

\section*{Appendix C: Particle-hole propagator and Macke function $I(q)$}
\setcounter{equation}{0}
\renewcommand{\theequation}{C.\arabic{equation}}

\noindent
{\bf C1:} The building elements of the RPA Feynman diagrams (cf Figs.2a, 3a, 4a, 5a) are the Coulomb repulsion
$v(q)= q_c^2/q^2$ with the coupling constant $q_c^2=4\alpha r_s/\pi$ and the
one-body Green's function of free electrons with $t(\vek k)= k^2/2$,
\begin{equation}\label{C1}
G_0(k,\omega)=\frac{\Theta(k-1)}{\omega-t(\vek k)+{\rm i}\delta}+
\frac{\Theta(1-k)}{\omega-t(\vek k)-{\rm i}\delta}\ , \quad  {\mbox{ $\delta{_> \atop ^{\to}}0$}}\ .
\end{equation}
{\bf C2:} From $G_0(k,\omega)$ follows the particle-hole propagator $Q(q,\eta)$ in RPA according to
\begin{equation}\label{C2}
Q(q,\eta)=-\int\frac{d^3k}{4\pi}\int\frac{d\omega}{2\pi{\rm i}}\ G_0(k,\omega)
G_0(|{\mbox{\bm $k$}}+{\mbox{\bm $q$}}|,\omega+\eta)
\end{equation}
with the result
\begin{eqnarray}\label{C3}
Q(q,\eta)&=&\int \frac{d^3k}{4 \pi} \left[\frac{\Theta^+(\vek k,\vek q)}{\eta-\tau(\vek k,\vek q)-{\rm i}\delta}
-\frac{\Theta^-(\vek k,\vek q)}{\eta - \tau(\vek k,\vek q)+{\rm i}\delta} \right]\ ,     \\ 
\tau(\vek k,\vek q)&=&t(\vek k+\vek q)-t(\vek k)=q\left (k\zeta+\frac{q}{2}\right) , \quad \zeta=\cos\varangle (\vek k,\vek q)\ .  \nonumber
\end{eqnarray}
The denominators contain the excitation energy $\tau(\vek k,\vek q)$, to create a hole with $\vek k$ inside the Fermi sphere and a particle with $\vek k+\vek q$
outside the Fermi sphere. A complicated step function, defined by
 $\Theta^{\pm}=1$ for $k\gtrless 1, |\vek k+\vek q|\lessgtr 1$ and 0 otherwise, is a consequence of the Pauli principle.  
$R(q,u)=Q(q,{\rm i}qu)$ defines a real function. A generalization of (\ref{C3}) is 
\begin{equation}\label{C4}
{\hat Q_i(q,\eta)f(\vek k_i)}=\int\frac{d^3k_i}{4\pi}\left[\frac{\Theta^+_i }{\eta-\tau_i-{\rm i}\delta}-
\frac{\Theta^-_i }{\eta-\tau_i+{\rm i}\delta}\right]f(\vek k_i)\ , \quad \tau_i=\tau(\vek k_i,\vek q)\ ,
\end{equation}
defining the integral operator $\hat Q_i(q,\eta)$, $i=1,2$.  \\

\noindent
{\bf C3:} When calculating the CSF $C_{\rm 1d}$, then the Macke function $I(q)$ appears via
\begin{equation}\label{C5}
\int\frac{d\eta}{2\pi{\rm i}}\ Q^2(q,\eta)=\frac{2}{(4\pi)^2}\ I(q)\ , \quad I(q)=
\int\frac{d^3k_1d^3k_2\Theta_{1,2}^-}{\vek q\cdot(\vek k_1+\vek k_2+\vek q)}\ .
\end{equation}
When dealing with the corresponding exchange term $C_{\rm 1x}$, then by means of contour integration it results
\begin{equation}\label{C6}
\int\frac{d\eta}{2\pi{\rm i}}{\hat Q_1(q',\eta)}{\hat Q_2(q',-\eta)v(q')|_{\vek q'=\vek k_1+\vek k_2+\vek q}}=
\frac{2}{(4\pi)^2}{\hat I(q)}v(|\vek k_1+\vek k_2+\vek q|)\ ,
\end{equation}
where the integral operator $\hat I(q)$ and the exchange term $I_{\rm x}(q)$ are defined by
\begin{eqnarray}
\hat I(q)v(\vek k_1+\vek k_2+\vek q)=\int\frac{d^3k_1d^3k_2\Theta_{1,2}^-}{\vek q\cdot(\vek k_1+\vek k_2+\vek q)}v(|\vek k_1+\vek k_2+\vek q|)=I_{\rm x}(q)v(q)\ .
\end{eqnarray} 
$I(q)$ and $I_{\rm x}(q)$ have the large-$q$ asymptotics
\begin{equation}\label{C8}
\frac{I(q\to \infty)}{(4\pi/3)^2}=\left(\frac{1}{q^2}+\frac{2}{5}\frac{1}{q^4}+\cdots\right)\ , \quad 
\frac{I_{\rm x}(q\to \infty)}{(4\pi/3)^2}=\left(\frac{1}{q^2}+\frac{x}{q^4}+\cdots\right)\ .
\end{equation}
Whether the guessed value of $x=8/5$ [text after (\ref{B7})] is correct, has to be studied. The change of 2/5 to 8/5 is a heavy enhancement of the $1/q^4$-tail, which is plausible
in view of the following qualitative arguments. In \cite{Zie5} and \cite{Zie11}, the (explicitly not known) function $I_{\rm x}(q)$ is compared with the explicitly known Macke function $I(q)$. Whereas $I(q)$ starts 
linearly with $q$ (what causes the Heisenberg
divergence), its exchange pendant $I_{\rm x}(q)$ must start at least with $q^2$. Otherwise
the integral $\int\limits_0^\infty dq\ I_{\rm x}(q)/q^2$ (which is proportional to the congenially calculated Onsager-et-al energy $v_{\rm x2}$) would not exist. An estimate shows even 
$I_{\rm x}(q\to 0)\sim q^3$. This flattening in the small-$q$ region together with the "area conservation" $\int\limits_0^\infty dq\ [I(q)-I_{\rm x}(q)]=0$ enforces the mentioned tail enhancement. \\ 

\noindent
{\bf C4:} Proof of (\ref{c25}): In the following the notation 
\begin{equation}\label{C9}
\Theta^+(\vek k,\vek q)=\Theta(k-1)\Theta(1-|\vek k+\vek q|), \quad \Theta^-(\vek k,\vek q)=\Theta(1-k)\Theta(|\vek k+\vek q|-1)
\end{equation}
is used.  (\ref{C4}) inserted in (\ref{c25}) gives 
\begin{eqnarray}\label{C10}
X(q,\eta)=
\left[\frac{\Theta(k_1-1)\Theta(1-|\vek k_1+\vek q|)}{\eta-\tau(\vek k_1,\vek q)-{\rm i}\delta}-\frac{\Theta(1-k_1)\Theta(|\vek k_1+\vek q|-1}{\eta-\tau(\vek k_1,\vek q)+{\rm i}\delta}\right]_{\vek k_1\to\vek k-\vek q}
\cdot \nonumber \\
\cdot\left[\frac{\Theta(k_2-1)\Theta(1-|\vek k_2+\vek q|)}{-\eta-\tau(\vek k_2,\vek q)-{\rm i}\delta}-\frac{\Theta(1-k_2)\Theta(|\vek k_2+\vek q|-1}{-\eta-\tau(\vek k_2,\vek q)+{\rm i}\delta}\right]_{\vek k_2\to-\vek k} \nonumber \\
=\left[\frac{\Theta(|\vek k-\vek q|-1)\Theta(1-k)}{\eta-\tau(\vek |\vek k-\vek q|,\vek q)-{\rm i}\delta}-\frac{\Theta(1-|\vek k-\vek q|)\Theta( k-1}{\eta-\tau(\vek |\vek k-\vek q|,\vek q)+{\rm i}\delta}\right]
\cdot \hspace{2cm}\nonumber \\
\cdot\left[\frac{\Theta(k-1)\Theta(1-|\vek k-\vek q|)}{-\eta-\tau(\vek k,-\vek q)-{\rm i}\delta}-\frac{\Theta(1-k)\Theta(|\vek k-\vek q|-1}{-\eta-\tau(\vek k,-\vek q)+{\rm i}\delta}\right]\ . \hspace{1.5cm} 
\end{eqnarray}
Only the products "over cross" contribute. With the substitution $\vek q\to -\vek q$ it results
\begin{equation}\label{11}
X(q,\eta)= \frac{\Theta(k-1)\Theta(1-|\vek k+\vek q|)}{[\eta +\tau(\vek k,\vek q)]^2}+\frac{\Theta(1-k)\Theta(|\vek k+\vek q|-1)}{[\eta +\tau(\vek k,\vek q)]^2}=
\frac{\tilde \Theta(\vek k,\vek q)}{[\eta+\tau(\vek k, \vek q)]^2}\ .
\end{equation}
To have finally exactly the same expression as in \cite{Zie11} the substitution $\vek k+\vek q\to -\vek k'$ changes the sign in the dominator, proving thus (\ref{c25}). \\

\noindent
{\bf C5:} Proof of (\ref{d1}): The diagram of Fig.5a means in detail
\begin{eqnarray}\label{C12}
\chi_{\rm lx2}(\vek r_1|\vek r'_1, \vek r_2|\vek r'_2)= \frac{1}{4}\int\frac{d^3q_1d^3q_2}{(4\pi/3)^2}\ \frac{\omega_{\rm pl}^4}{q_1^2q_2^2}\int\frac{d^3k_1d^3k_2}{(4\pi/3)^2} \cdot \nonumber \\
\cdot\frac{1}{2}\left\{{\rm e}^{{\rm i}[(\vek k_2+\vek q_1+\vek q_2)\vek r_{12'}+\vek k_1\vek r_{21'}+(\vek q_1+\vek q_2)\vek r_{12}]}\ \chi_{\rm lx2}(\vek k_{1,2},\vek q_{1,2})+{\rm h.c.}\right\}\ ,
\end{eqnarray}
where the following abbreviations are used:
\begin{equation}\label{C13}
\chi_{\rm lx2}(\vek k_{1,2},\vek q_{1,2})= \int\frac{d\omega_1d\omega_2d\eta_1d\eta_2}{(2\pi{\rm i})^4}\cdot x_1(\vek k_1)\cdot x_2(\vek k_2)\quad {\rm with}
\end{equation}
\begin{equation}
x_1(\vek k_1)=\frac{(k_1\gtrless 1)}{\omega_1-t(\vek k_1)\pm{\rm i}\delta}\ \cdot 
\frac{(|\vek k_1+\vek q_1|\gtrless 1)}{\omega_1+\eta_1-t(\vek k_1+\vek q_1)\pm{\rm i}\delta}\ \cdot\frac{(|\vek k_1+\vek q_1+\vek q_2|\gtrless 1)}{\omega_1+\eta_1-\eta_2-t(\vek k_1+\vek q_1+\vek q_2)\pm{\rm i}\delta}\ , \nonumber 
\end{equation}
\begin{equation}
x_2(\vek k_2)= \frac{(k_2\gtrless 1)}{\omega_2-t(\vek k_2)\pm{\rm i}\delta}\ \cdot 
\frac{(|\vek k_2+\vek q_2|\gtrless 1)}{\omega_2-\eta_2-t(\vek k_2+\vek q_2)\pm{\rm i}\delta}\ \cdot\frac{(|\vek k_2+\vek q_1+\vek q_2|\gtrless 1)}{\omega_2+\eta_1-\eta_2-t(\vek k_2+\vek q_1+\vek q_2)\pm{\rm i}\delta}
\ . \nonumber  
\end{equation}
For the contribution of the ladder diagram $\chi_{\rm lx2}$ to the MD $n(k)$, 
 the contraction steps analog (\ref{c21}) can be performed with the result
\begin{equation}
\int\frac{d^3r_{11'}}{2\cdot 3\pi^2}{\rm e}^{-{\rm i}\vek k \vek r_{11'}}\ \int\frac{d^3r_2}{2\cdot 3\pi^2}{\rm e}^{{\rm i}[(\vek k_2+\vek q_1+\vek q_2)\vek r_{12}+\vek k_1\vek r_{21'}+(\vek q_1+\vek q_2)\vek r_{12}]}
=\left(\frac{4\pi}{3}\right)^2 \delta(\vek k_1-\vek k)\delta(\vek k_2-\vek k_1) \nonumber
\end{equation}
\begin{equation}\label{C14}
\curvearrowright \quad {\rm \textroundcap{$\chi$}}_{\rm lx2}(k)= \frac{1}{4}\int\frac{d^3q_1d^3q_2}{(4\pi/3)^2}\ \frac{\omega_{\rm pl}^4}{q_1^2q_2^2}\frac{1}{2}\left\{\chi_{\rm lx2}(\vek k,\vek q_{1,2})+{\rm c.c.}
\right\}\ . 
\end{equation}
$\chi_{\rm lx2}(\vek k,\vek q_{1,2})$ contains contour integrations in 4 complex frequency planes. Unfortunately, each of these 4 integrals runs over 3 frequency denominators in a difficult way. 
But by means of a trick ( integration variables are substituted) the $\eta_{1,2}$ integrations can be decoupled: $\omega_2\to \omega_2-\eta_1+\eta_2$. This does not influence $x_1(\vek k)$ (it is only written in 
reverse order), but it changes $x_2(\vek k)$:
\begin{equation}\label{C15}
x_1(\vek k_1)=\frac{(|\vek k_1+\vek q_1+\vek q_2|\gtrless 1)}{\omega_1+\eta_1-\eta_2-t(\vek k_1+\vek q_1+\vek q_2)\pm{\rm i}\delta}\cdot
\frac{(|\vek k_1+\vek q_1|\gtrless 1)}{\omega_1+\eta_1-t(\vek k_1+\vek q_1)\pm{\rm i}\delta}\ \cdot 
\frac{(k_1\gtrless 1)}{\omega_1-t(\vek k_1)\pm{\rm i}\delta}\ ,  
\end{equation}
\begin{equation}\label{C16}
x_2(\vek k)= \frac{k\gtrless 1}{\omega_2-\eta_1+\eta_2-t(\vek k)\pm{\rm i}\delta}\cdot 
\frac{|\vek k+\vek q_2|\gtrless 1}{\omega_2-\eta_1-t(\vek k+\vek q_2)\pm{\rm i}\delta}\cdot
\frac{|\vek k+\vek q_1+\vek q_2|\gtrless 1}{\omega_2-t(\vek k+\vek q_1+\vek q_2)\pm{\rm i}\delta}\ .
\end{equation}
Now it is easy to perform the integration over $\eta=\eta_2-\eta_1$, namely
\begin{equation}\label{C17}
\int \frac{d\eta}{2\pi{\rm i}}\cdot
\frac{(|\vek k+\vek q_1+\vek q_2|\gtrless 1)}{\omega_1-\eta-t(\vek k+\vek q_1+\vek q_2)\pm{\rm i}\delta_1}\cdot
\frac{(k\gtrless 1)}{\omega_2+\eta-t(k)\pm{\rm i}\delta_2}=
\frac{(\mp)(k\gtrless 1,|\vek k+\vek q_1+\vek q_2|\gtrless 1)}{\omega_1+\omega_2-t(\vek k)-t(\vek k+\vek q_1+\vek q_2)\pm{\rm i}\delta}\ .
\end{equation}
Therein those cases do not contribute, where the 2 poles in the complex $\eta$-plane are both on the same side of the real axis. Only the combinations $(\pm {\rm i}\delta_1)$ with $(\mp {\rm i}\delta_2)$ contribute.
Similarly the $\eta_1$-integration is handled:
\begin{equation}\label{C18}
\int \frac{d\eta_1}{2\pi{\rm i}}\cdot
\frac{(|\vek k+\vek q_1|\gtrless 1)}{\omega_1+\eta_1-t(\vek k+\vek q_1)\pm{\rm i}\delta_1}\cdot
\frac{(|\vek k+\vek q_2|\gtrless 1)}{\omega_2-\eta_1-t(\vek k+\vek q_2)\pm{\rm i}\delta_2}=
\frac{(\mp)(|\vek k+\vek q_{1,2}|\gtrless 1)}{\omega_1+\omega_2-t(\vek k+\vek q_1)-t(\vek k+\vek q_2)\pm{\rm i}\delta}\ .
\end{equation}
\noindent
The next step would be the $\omega_{2}$-integration. The integrand is a product of the rhs's of (\ref{C17}) and (\ref{C18}) with the 3rd factor of (\ref{C16}):
\begin{eqnarray}\label{C19}
\int\frac{d\omega_2}{2\pi{\rm i}}\cdot\frac{(\mp)(k\gtrless 1,|\vek k+\vek q_1+\vek q_2|\gtrless 1)}{\omega_1+\omega_2-t(\vek k)-t(\vek k+\vek q_1+\vek q_2)\pm{\rm i}\delta}&\cdot& \nonumber \\
\cdot\ \frac{(\mp)(k\gtrless 1,|\vek k+\vek q_{1,2}|\gtrless 1)}{\omega_1+\omega_2-t(\vek k)-t(\vek k+\vek q_1)-t(\vek k+\vek q_2)\pm{\rm i}\delta}&\cdot&  \\
\cdot\ \frac{(|\vek k+\vek q_1+\vek q_2|\gtrless 1)}{\omega_2-t(\vek k+\vek q_1+\vek q_2)\pm{\rm i}\delta}&& =\
\frac{\Theta^\pm(\vek k,\vek q_{1,2})}{[\omega_1-t(\vek k)+\vek q_1\cdot\vek q_2\mp{\rm i}\delta]\ \vek q_1\cdot\vek q_2}\ . \nonumber
\end{eqnarray}    
The integration region is defined by the functions 
$\Theta^{\pm}(\vek k,\vek q_{1,2})=1 \quad {\rm for} \quad (k\gtrless 1, |\vek k+\vek q_1+\vek q_2|\gtrless 1,|\vek k+\vek q_{1,2}| \lessgtr 1) \quad {\rm and}\quad 0 \quad {\rm otherwise}$. The identity 
\begin{equation}\label{C20}
t(\vek k)+t(\vek k+\vek q_1+\vek q_2)-t(\vek k+\vek q_1)-t(\vek k+\vek q_2)=\vek q_1\cdot\vek q_2
\end{equation}
is used. The remaining $\omega_1$-integration finally yields
\begin{equation}\label{C21}
\chi_{\rm lx2}(\vek k,\vek q_{1,2})=\int\frac{d\omega_1}{2\pi{\rm i}}\frac{\Theta^\pm(\vek k,\vek q_{1,2})}{[\omega_1-t(\vek k)\pm{\rm i}\delta][\omega_1-t(\vek k)+\vek q_1\cdot \vek q_2\mp{\rm i}\delta]\ \vek q_1\cdot\vek q_2}=
\frac{\Theta(\vek k,\vek q_{1,2})}{(\vek q_1\cdot\vek q_2)^2}
\end{equation}
where $\Theta=\Theta^++\Theta^-$. (\ref{C21}) inserted in (\ref{C14}) gives (\ref{d1}), qed .

\section*{Appendix D: Correlation parameters}
\setcounter{equation}{0}
\renewcommand{\theequation}{D.\arabic{equation}}

\noindent
Here is a list of characteristic correlation parameters as functions of $r_s$, vanishing for $r_s\to 0$: \\ 
$N=\int_0^\infty d(k^3)\ [n(k)-n_0(k)]$ is the normalization of the MD-correlation tails. \\
$c=\int_0^\infty d(k^3)\ n(k)[1-n(k)]=N-\int_0^\infty d(k^3)\ [ n(k)-n_0(k)]^2$ is the L\"owdin parameter.  \\
$1-z_{\rm F}$ is the reduced discontinuity jump of $n(k)$ at the Fermi surface $|\vek k|=1$. \\
$s=-\int_0^\infty d(k^3)\ [n(k)\ln n(k)+(1-n(k))\ln (1-n(k))]$ is the particle-hole symmetric correlation entropy, see Eq.(22) in \cite{GGZ}. \\
$c_1=h_{\rm a}(0)$ with  $1-h_{\rm a}(0)=g_{\rm a}(0)$ being the on-top value of the Coulomb hole, \\ 
$c_1$ simultaneously determines the large-$q$ asymptotics of $C_{\rm d,x}(q)$ and $n(k)$, \\
$c_{2},c_{2}'$ determine the higher-order large-$q$ asymptotics of $C_{\rm d,x}(q)$, \\  
$c_3=h_{\rm p}''(0)$ and $t$ determine the on-top curvature of the Fermi hole $g_{\rm p}''(0)=\frac{4}{3}t-h_{\rm p}''(0)$, \\
$c_{4,5}$ determine the small-$q$ behavior of $S(q)$ beyond the plasmon-SR term $q^2/(2\omega_{\rm pl})$, \\ 
$c_6$ influences the jump discontinuity $\Delta S''(2)$.\\
All these parameters may serve as measures of the correlation strength. They vanish for $r_s\to 0$ with terms $\sim r_s$, $r_s^2\ln r_s$ or $r_s^{3/2}$. 
From a diagramatical point of view all the quantities characterizing $n(k)$ or derived from it, should start with $r_s^2$. The reason: the diagrams for $n(k)$ start with $n_2(k)$, the 1st-order term $n_1(k)$
(Fig. 4c) is zero - in agreement with the vanishing 1st-order kinetic energy $t_1=0$, what follows from the virial theorem (\ref{a1}). But there are starting dependencies as $N=O(r_s^{3/2})$ cf (3.46) in 
\cite{Zie11}. This peculiar behavior is an example for how higher-order 
partial summations may create lower-order terms. It is a peculiarity of the Coulomb repulsion. $N\sim r_s^{3/2}$ results from that $k$-region $|k-1|\ll q_c$, where the RPA reconstruction takes place.   
A characteristic non-analyticity is also the behavior of $n(k)$ at the Fermi surface $|\vek k|=1$, namely $\sim (k-1)\ln |k-1|$.
\end{appendix}

\clearpage

\begin{center}
{\bf Figures}
\end{center}

\begin{figure}[h!]
\begin{center}
\includegraphics[scale=0.5]{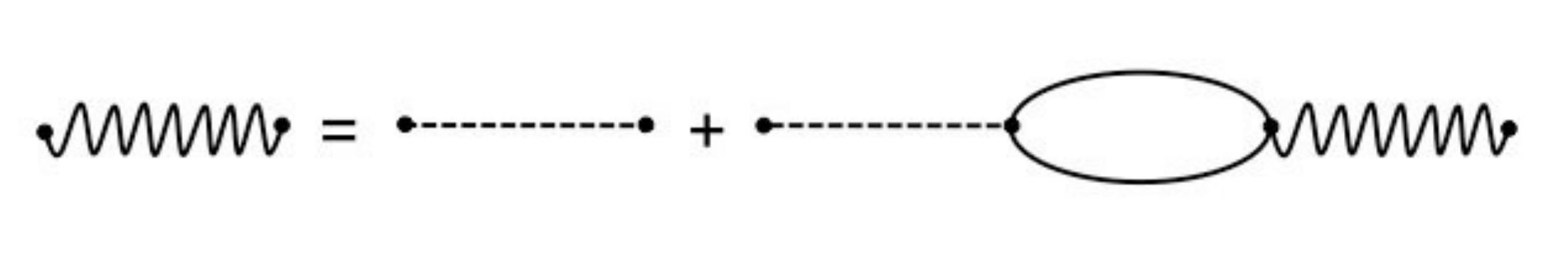}
\end{center}
\caption{Feynman diagrams for the Coulomb repulsion in RPA with the partial summation (or screening replacement) $v(q)\to v(q,\eta)$, $q$ = momentum transfer, $\eta$ = energy/frequency, dashed line = bare interaction $v(q)=q_{\rm c}^2/q^2$,
closed loop = particle-hole propagator
$Q(q,\eta)$ of (\ref{C3}),  wavy line = effectively screened interaction $v(q,\eta)=v(q)/[1+v(q)Q(q,\eta)]$.}
\end{figure}

\begin{figure}[h!]
\begin{center}
\includegraphics[scale=0.5]{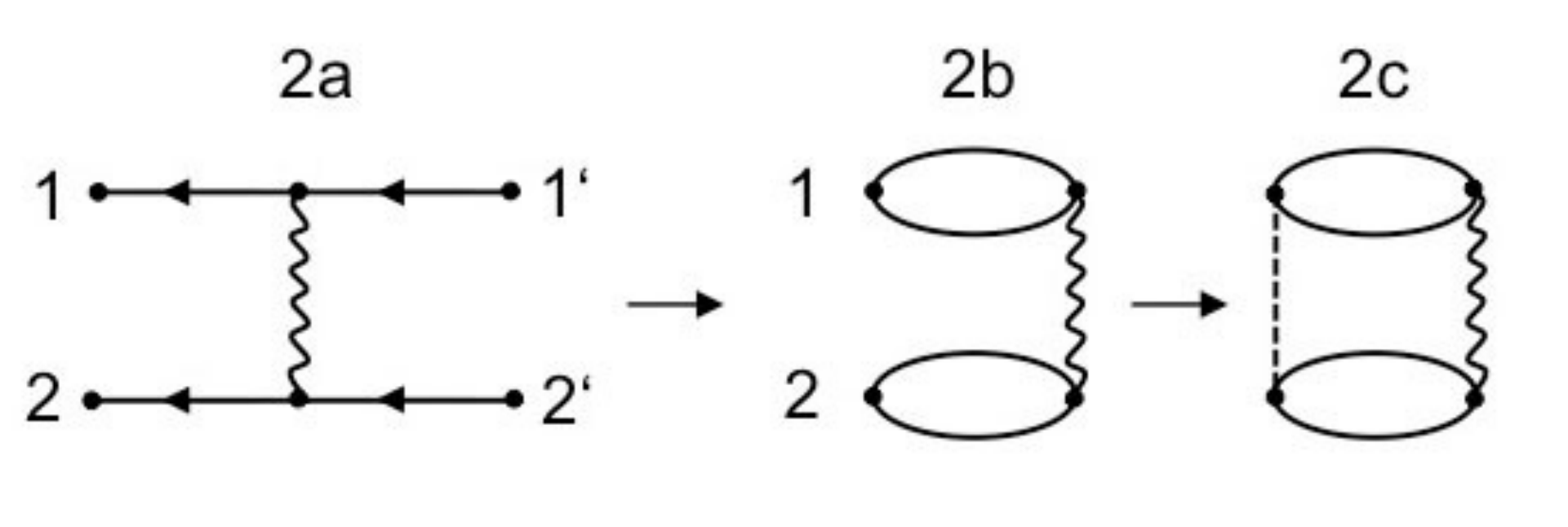}
\end{center}
\caption{2a = lowest-order renormalized cumulant 2-matrix $\chi_{\rm dr}$, 2b = non-divergent cumulant PD $h_{\rm dr}$ or cumulant SF $C_{\rm dr}$ (Kimball) \cite{Kim1}, 2c = non-divergent interaction energy $v_{\rm dr}$ (Macke, Gell-Mann/Brueckner) \cite{Ma,GB}.}
\end{figure}

\begin{figure}[h!]
\begin{center}
\includegraphics[scale=0.5]{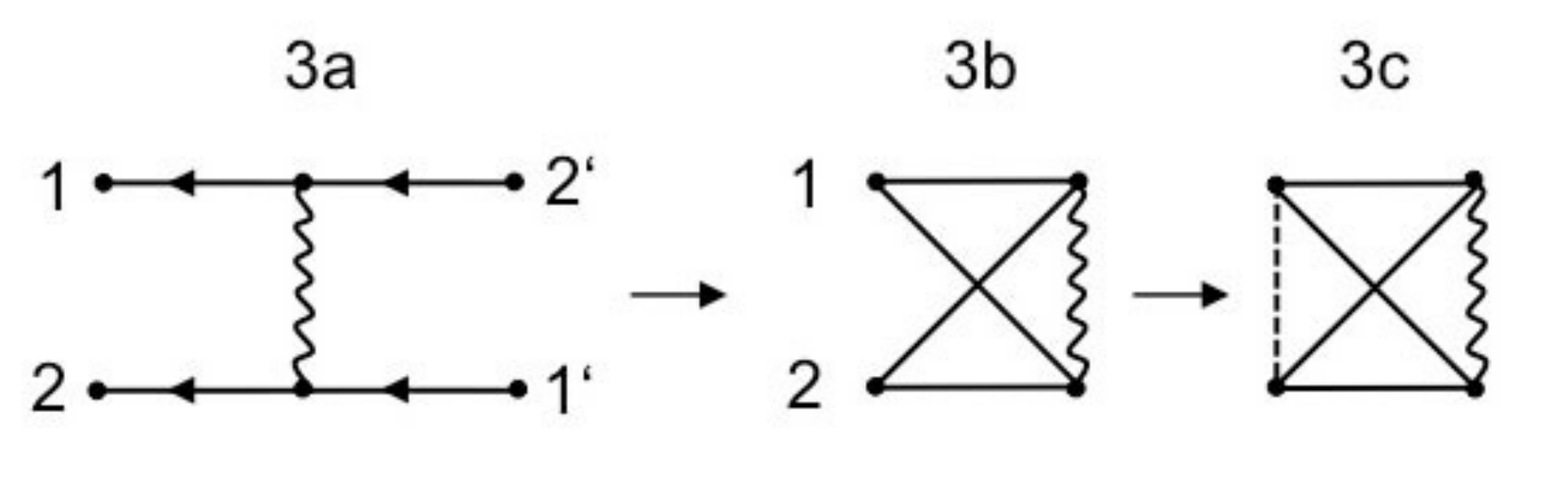}
\end{center}
\caption{The RPA exchange terms corresponding to Fig.2 with 3a = $\chi_{\rm xr}$ (following from 2a = $\chi_{\rm dr}$ through the exchange replacement
$1'\leftrightarrow 2')$, 3b = $h_{\rm xr}$ or $C_{\rm xr}$ (non-divergent), and 3c = $v_{\rm xr}$. As shown by Onsager et al. \cite{Ons}, already $v_{\rm x2}$ does not diverge (needs no RPA renormalization).}
\end{figure}

\newpage

\begin{figure}[h!]
\begin{center}
\includegraphics[scale=0.6]{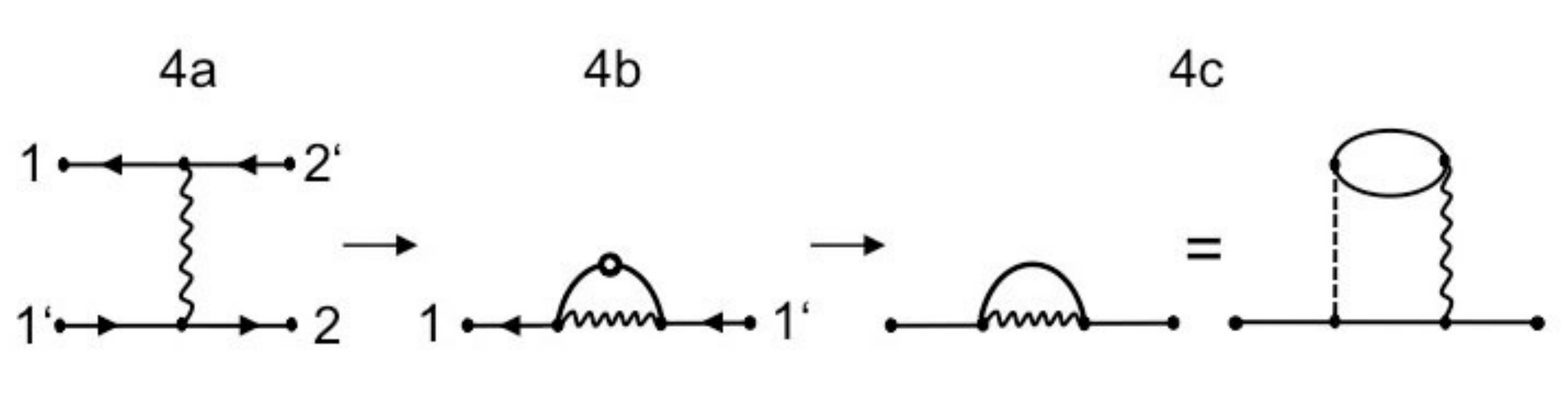}
\end{center}
\caption{How the 2-body diagram 4a = 3a = $\chi_{\rm xr}$ transforms to the 1-body diagram 4b through the contraction [with 2'=2 and $\int d2$, see (\ref{b25})], marked by a small circle   and
finally to the non-divergent RPA-diagram $n_{\rm r}(k)$ of Fig.4c (Daniel/Vosko and Kulik) \cite{Da,Ku}. The 1st-order term is zero (analog to the 1st-order kinetic energy $t_1=0$).} 
\end{figure}

\begin{figure}[h!]
\begin{center}
\includegraphics[scale=0.6]{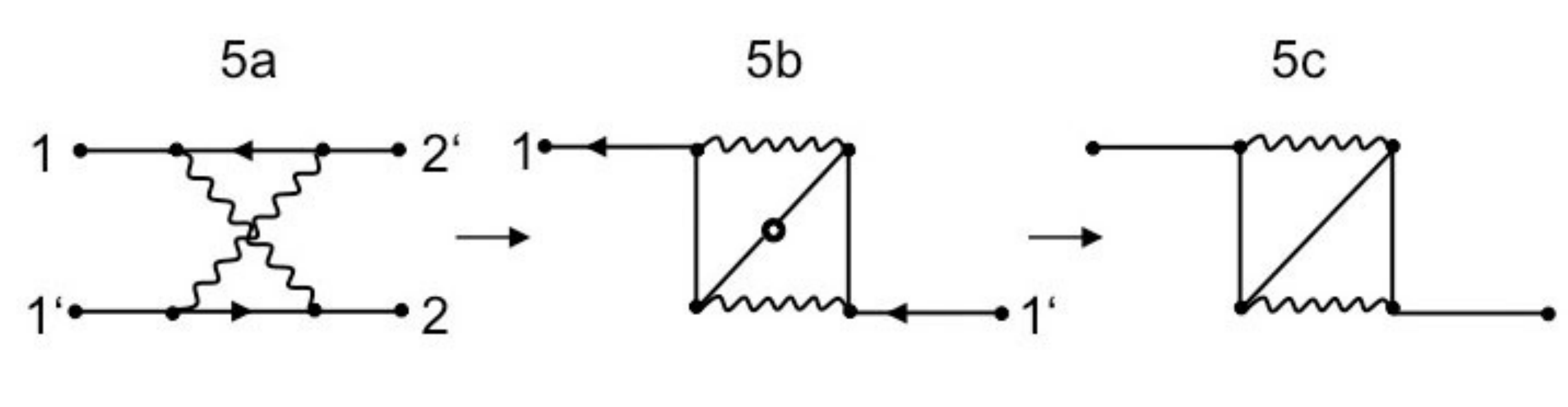}
\end{center}
\caption{How the 2-body ladder diagram 5a in its exchange version $\chi_{\rm lxr}$ transforms to the 1-body diagram 5b through the contraction (\ref{b25}) and finally to the diagram $n_{\rm x}(k)$ of Fig.5c.
In Sec.IV and App.C5 the RPA screening is neglected for simplicity corresponding to the replacement r$\to$2 or $v(q,\eta)\to v(q)$.} 
\end{figure}

\begin{figure}[h!]
\begin{center}
\includegraphics[scale=0.6]{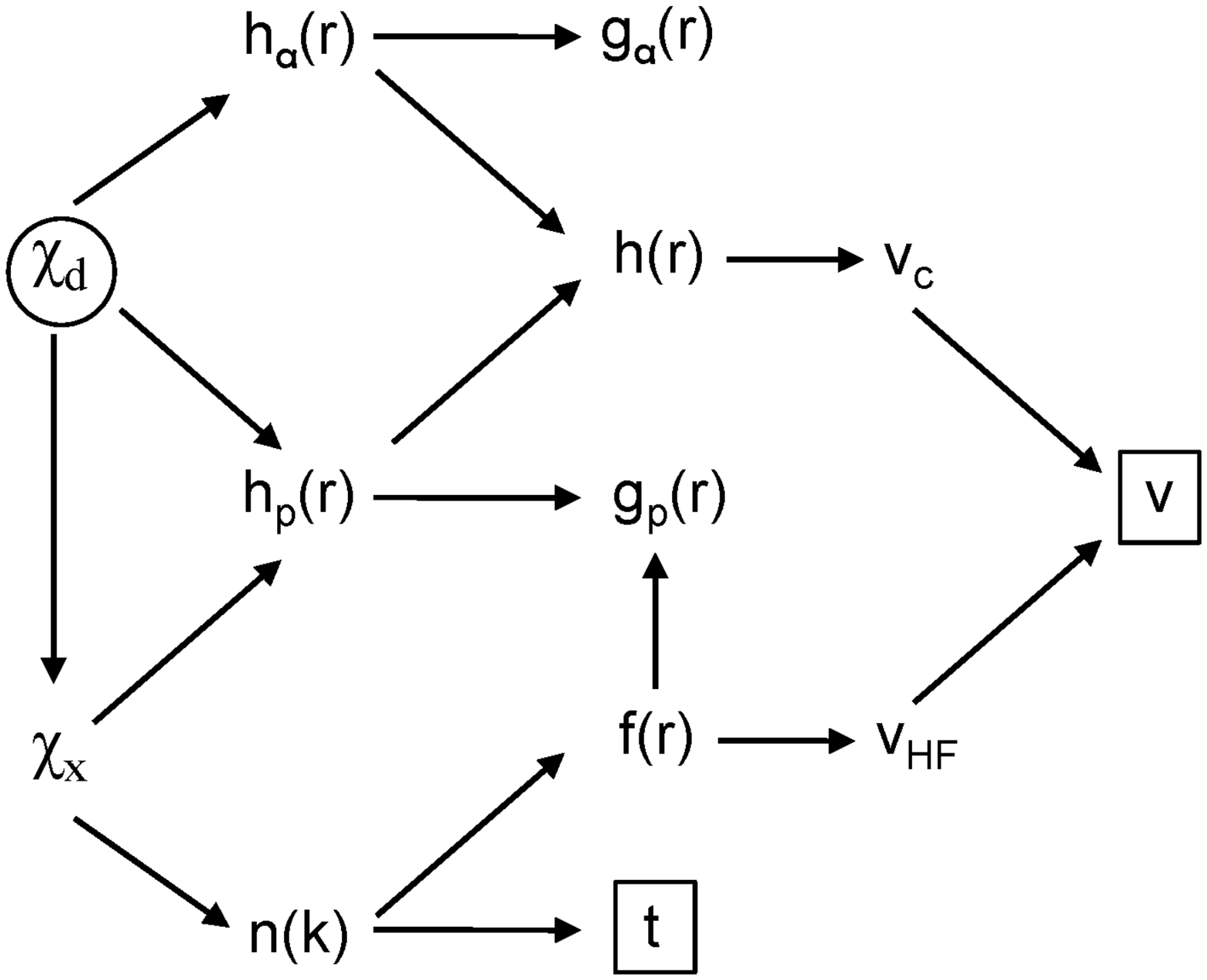}
\end{center}
\caption{Flow charts showing the way from the starting point $\chi_{\rm d}$ = direct cumulant 2-matrix through its exchange pendant $\chi_{\rm x}$ and the cumulant PD $h(r)$ to the final results $n(k)$ and $g(r)$,
from which follow (i) $f(r)$ = 1-matrix,
$t$ = kinetic energy, $c$ = L\"owdin parameter and (ii) $v$ = interaction energy, respectively. "a"=spin-antiparallel, "p"=spin-parallel. From FT arises a similar scheme for the structure factor $S(q)$, the cumulant
structure factor $C(q)$, and the HF term $F(q)$ [$\leftrightarrow g(r),\ h(r),\ {\rm and}\ f^2(r)$, respectively]. Note that $g(r)=\frac{1}{2}[g_{\rm a}+g_{\rm p}(r)]$ and $e=t+v$.}
\end{figure}

\newpage

\clearpage
\includegraphics[angle=90]{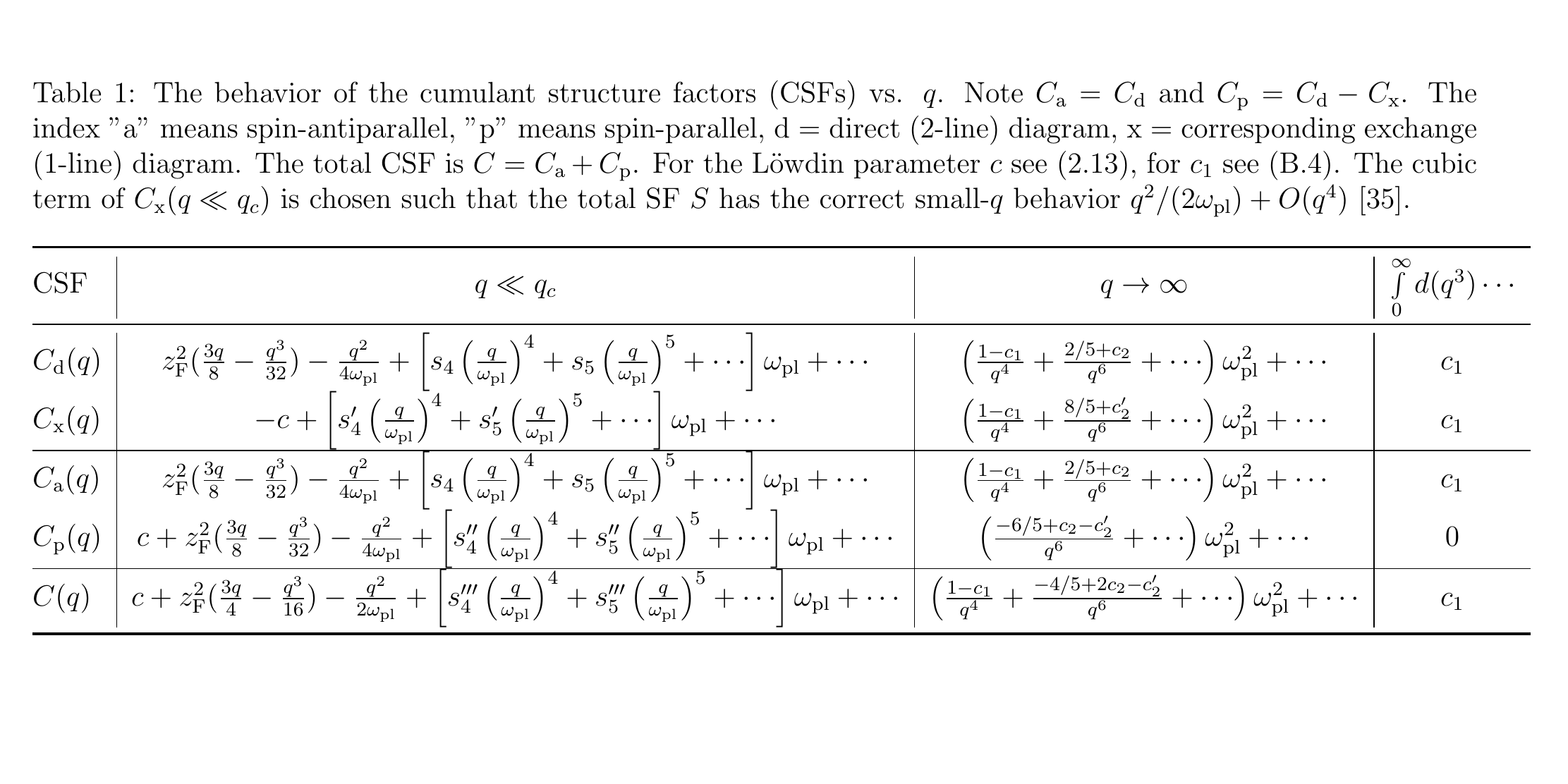}

\clearpage
\includegraphics[angle=90]{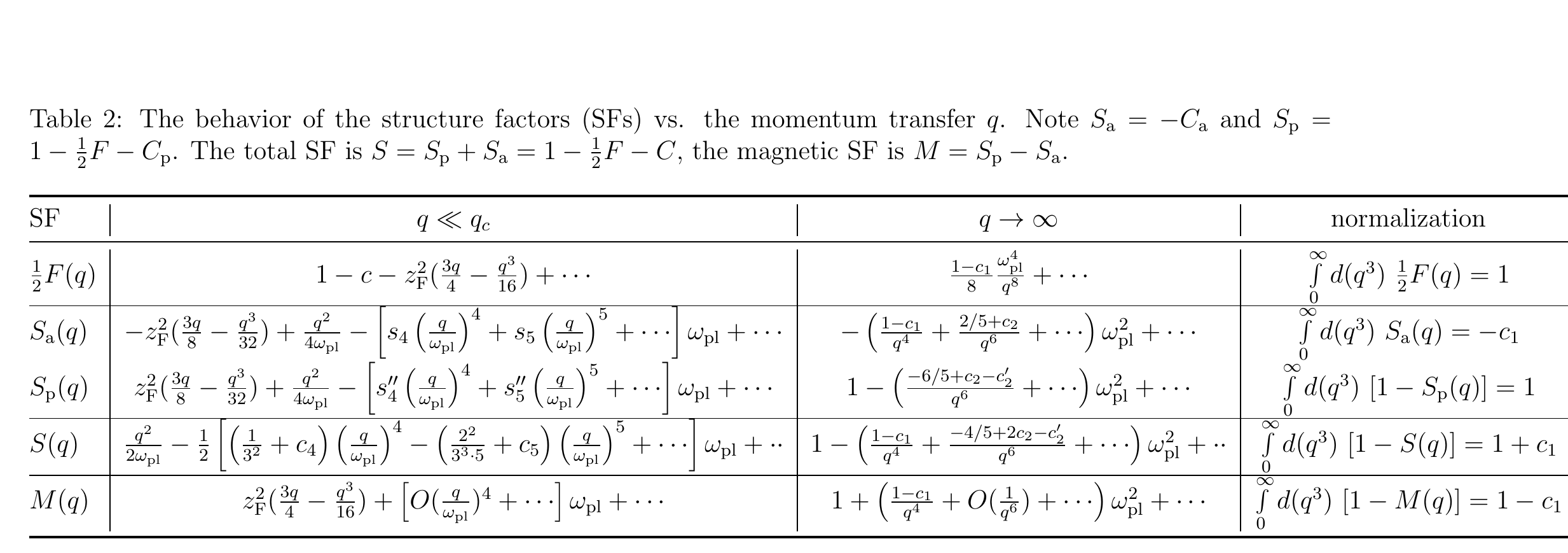}


\begin{thebibliography}{1}
\bibitem{foo}  The Coulomb repulsion takes the form $\alpha r_s/r$, if lengths and energies are measured in units of
$k_{\rm F}^{-1}$ and $k_{\rm F}^2$, respectively. This shows, that $r_s$ is not only a density parameter \cite{Zie1},
but also measures the interaction strength or coupling constant. The FT of $\alpha r_s/r$ is 
$4\pi\alpha r_s/q^2=\pi^2(q_c^2/q^2)$ with $q$ measured in units of $k_{\rm F}$ and $q_c^2=4\alpha r_s/\pi$. - Notice that $\tilde e=k_{\rm F}^2e=e/(\alpha r_s)^2$ gives the energy
in a.u., e.g. the energy in zeroth order and the lowest-order exchange energy are ${\tilde e}_0=3/(10\ \alpha^2 r_s^2)$
and ${\tilde e}_{\rm x}=-3/(4 \pi \alpha r_s)$, respectively.
\bibitem{foo1} The GS-energy $e$ is also a functional of $t(\vek k)$ and $v(\vek q)$ with $\delta e/\delta t(\vek k)\sim n(k)$ \cite{Da} and $\delta e/\delta v(\vek q)\sim S(q)-1$ \cite{Yas,Kim3}.
\end{thebibliography}
\end{document}